\documentclass[fleqn,usenatbib]{mnras}

\usepackage{newtxtext,newtxmath}
\usepackage{gensymb}

\usepackage[T1]{fontenc}

\DeclareRobustCommand{\VAN}[3]{#2}
\let\VANthebibliography\thebibliography
\def\thebibliography{\DeclareRobustCommand{\VAN}[3]{##3}\VANthebibliography}

\usepackage{graphicx}	
\usepackage{amsmath}	
\usepackage{multirow}
\usepackage{orcidlink}      
\usepackage{ulem}

\DeclareMathOperator{\sech}{sech}

\title[NS+WD binaries with  LISA]{Neutron Star - White Dwarf Binaries: Probing Formation Pathways and Natal Kicks with LISA}

\author[V. Korol et al.]{Valeriya Korol$^{\orcidlink{0000-0002-6725-5935}}$$^{1,2}$\thanks{E-mail: korol@mpa-garching.mpg.de (VK)},
Andrei P. Igoshev$^{\orcidlink{0000-0003-2145-1022}}$$^{3}$,
Silvia Toonen$^{\orcidlink{0000-0002-2998-7940}}$$^{4}$,
Nikolaos Karnesis$^{\orcidlink{0000-0002-2380-3186}}$$^{5}$,
Christopher J. Moore$^{\orcidlink{0000-0002-2527-0213}}$$^{2}$,
\newauthor 
Eliot Finch$^{\orcidlink{0000-0002-1993-4263}}$$^{6,2}$, and
Antoine Klein$^{\orcidlink{0000-0001-5438-9152}}$$^{2}$ 
\\
$^{1}$Max-Planck-Institut f{\"u}r Astrophysik, Karl-Schwarzschild-Stra{\ss}e 1, 85748 Garching, Germany\\
$^{2}$Institute for Gravitational Wave Astronomy \& School of Physics and Astronomy, University of Birmingham, Birmingham, B15 2TT, UK\\
$^{3}$Department of Applied Mathematics, University of Leeds, LS2 9JT Leeds, UK\\
$^{4}$Anton Pannekoek Institute for Astronomy, University of Amsterdam, 1090 GE Amsterdam, The Netherlands\\
$^{5}$Department of Physics, Aristotle University of Thessaloniki, Thessaloniki 54124, Greece\\
$^{6}$TAPIR, California Institute of Technology, Pasadena, CA 91125, USA\\
}

\date{Accepted 2024 March 22. Received 2024 March 22; in original form 2023 October 10}

\pubyear{2023}

\begin{document}
\label{firstpage}
\pagerange{\pageref{firstpage}--\pageref{lastpage}}
\maketitle

\begin{abstract}
Neutron star-white dwarf (NS+WD) binaries offer a unique opportunity for studying NS-specific phenomena with gravitational waves. In this paper, we employ the binary population synthesis technique to study the Galactic population of NS+WD binaries with the future Laser Interferometer Space Antenna ({\it LISA}). We anticipate approximately $\mathcal{O}(10^2)$ detectable NS+WD binaries by {\it LISA}, encompassing both circular and eccentric ones formed via different pathways. Despite the challenge of distinguishing these binaries from more prevalent double white dwarfs (especially at frequencies below 2\,mHz), we show that their eccentricity and chirp mass distributions may provide avenues to explore the NS natal kicks and common envelope evolution. Additionally, we investigate the spatial distribution of detectable NS+WD binaries relative to the Galactic plane and discuss prospects for identifying electromagnetic counterparts at radio wavelengths. Our results emphasise {\it LISA}'s capability to detect and characterise NS+WD binaries and to offer insights into the properties of the underlying population. Our conclusions carry significant implications for shaping {\it LISA} data analysis strategies and future data interpretation.
\end{abstract}




\section{Introduction}

Neutron star - white dwarf (NS+WD) binaries hold significant potential in advancing our understanding of binary neutron star (NS) formation channels, natal kicks, and other NS-specific physics. In recent years, the field has experienced considerable growth particularly in the area of hydrodynamical and nuclear–hydrodynamical simulations, including the first 1-dimensional simulations of NS+WD mergers \citet{mez12}, followed by significant advancements in 2-dimensional simulations by \citet{fer13} and \citet{zen19}, and the introduction of 3-dimensional smooth particle hydrodynamic simulations in \citet{bob22} and magneto-hydrodynamic simulations in \citet{mor23}. These simulations have highlighted the importance of NS+WD mergers within the transient astronomy field, contributing to a deeper understanding of various transient phenomena \citep[see also][]{mar16,kat22,kan23}. Moreover, studies have also underscored the role of NS+WD binaries in the formation of ultra compact X-ray binaries \citep{tau18}, Thorne-\.{Z}ytkow objects-like objects \citep{pas09}, rocky planets around pulsars \citep{mag17}, and planetary nebulae \citep{abl24}. The forthcoming \textit{Laser Interferometer Space Antenna} (\textit{LISA}; \citealt{LISAproposal, LISA2Redbook}) -- as well as similar planned space-based gravitational wave (GW) observatories \textit{TianQin} \citep{TianQin, hua20} and \textit{Taiji} \citep{Taiji} -- offers a unique opportunity to discover NS+WD binaries and shed light on the underlying physics that govern their formation and evolution. In addition, in a more distant future, proposed GW missions like the \textit{Lunar Gravitational Wave Antenna} \citep{har21,bra23} and the \textit{DECi-hertz Interferometer Gravitational wave Observatory} \citep{DECIGO,arc20} will capture GW signals from NS+WD mergers at deci-Hertz frequencies \citep[][]{mor23}.

Although neutron star - neutron star (NS+NS) systems have received more attention in the context of the {\it LISA} mission \citep{yu15,kyu19,and20,lau20,kor21,wag22,set22,sto22}, it is becoming increasingly clear that NS+WD binaries may provide cleaner probes for studying NS natal kicks and other binary evolution associated phenomena \citep[e.g.][]{tau18,rui19,he24}. Firstly, NS+WD systems typically have a less complex evolutionary history, involving only one supernova during their formation. This simpler history makes it easier to disentangle the effects of natal kicks from other processes, allowing for a more direct assessment of the kicks' impact on the binary system. Furthermore, the relative rarity of NS+NS systems compared to NS+WD binaries means that there is a larger population of the latter available for study \citep[e.g.][]{nel01a,bre20}. By analysing a greater number of NS+WD systems, in addition to known populations of binary radio pulsars with WD companions observed and ultra-compact X-ray binaries \citep[e.g.][]{riv15,kru21,pad23}, we can obtain more robust statistics on their properties. Examining the NS+WD population's orbital characteristics, such as periods and eccentricities, can provide insights into their formation channels and the role of natal kicks \citep{LISAastro}. 

In this work, we employ a suite of NS+WD binary population models from \citet{too18}.
We focus on the NS+WD parameters measurable with {\it LISA}, including the chirp mass, eccentricity, and the 3-dimensional (3D) position in the Milky Way, seeking signatures that provide insights into NS natal kicks and NS+WD formation pathways. The chirp mass -- a combination of the component masses that determines the GW signal's evolution -- is a crucial parameter for distinguishing NS+WD from other types of Galactic binaries accessible to {\it LISA} \citep[mainly WD+WD and NS+NS, e.g., see][]{LISAastro}. In addition, we anticipate that the chirp mass can be used to distinguish between models involving different mass-transfer prescriptions. Eccentricity can also provide valuable information about the dynamical interactions and mass-transfer processes that shaped the system's orbital evolution. For example, \citet{lau20} have emphasised the importance of eccentricity in distinguishing various NS+NS formation channels, highlighting the crucial role of the last mass-transfer phase prior to the NS formation in determining system's characteristics. Finally, we also explore whether the observed Galactic distribution of NS+WD binaries can reveal insights into NS formation processes. Drawing from the methodology of \citet{Repetto2012MNRAS,rep17}, we also investigate if an offset of a NS+WD system from the Galactic plane can serve as a signature of peculiar velocity with respect to circular Galactic motion, expected for systems not receiving kicks. 

\section{Assembling a synthetic NS+WD population} \label{sec:methods}

We construct a representative present-day Galactic population of NS+WD binaries through a three-step process. Firstly, we utilise a suite of binary population synthesis models from \citet{too18} to account for the evolution from the initial main sequence (MS) stage to NS+WD formation (see Section~\ref{sec:MStoNS+WD}). 
Each simulated catalogue represents a stellar population of approximately $\sim10^8$~M$_\odot$, accounting for the fact that stars may be single or in binary systems, and may or may not have reached the main sequence turnoff within the Hubble time. These models provide binary properties at the time of NS+WD formation. Secondly, we seed these NS+WD binaries within a Milky Way potential at a rate determined by an adopted star formation history (see Section~\ref{sec:MSMStopresent}). From the NS+WD formation to the present day, we evolve the binaries' orbital parameters in accordance with GW radiation reaction (Section~\ref{sec:orb_evol}). Lastly, we assign to binaries' 3D positions in the Milky Way gravitational potential and integrate their orbits from the time of NS formation (when receiving the natal kick) until the present age of the Galaxy (Section~\ref{sec:MWmodel}). We provide a detailed description of these steps below.

\subsection{From MS+MS to NS+WD} \label{sec:MStoNS+WD}

We utilise a suite of nine NS+WD population models compiled in \citet{too18} using the {\sc SeBa} stellar and binary evolution module \citep{spz96,nel01,too12}. Below we summarise the set of the assumptions used to generate primordial binaries and refer the reader to \citet[][section~3.2]{too18} for further details and discussion. Firstly, masses of primaries ($m_1$) are drawn from the initial mass function of \citet[][see also \citealt{kro08}]{kro93} focusing on a mass range of $4-25$ M$_\odot$. We note that we consider a broader range from 0.1 to 100 M$_\odot$ to calculate the normalisation (i.e. the corresponding simulated stellar mass) of this population. Masses of secondaries ($m_2$) are then determined based on a flat mass ratio distribution where $0 < m_2/m_1 < 1$ \citep{rag10,duc13}. The orbital separations are drawn from a flat distribution in $\log(a)$, following \citet{abt83}, while initial eccentricities are derived from a thermal distribution, as per \citet{heg75}. Finally, we assume a constant binary fraction of 75\,per~cent, which is supported by observations in the considered mass range \citep{rag10,duc13,san14}. 
In each model, primordial binaries and most input physics remain consistent, with variations only in the NS natal kick and common envelope prescriptions (see Sections \ref{sec:kicks} and \ref{sec:CE}). The primary goal of this study is to understand differences in the {\it LISA}-detectable NS+WD population when varying the natal kick prescription to assess whether NS natal kicks can be constrained based on the {\it LISA} data. Variations in the  common envelope prescription aim to assess the uncertainty range in the {\it LISA}-detectable population, as common envelope prescriptions have been found to produce the largest differences in the population synthesis of compact binaries \citep[e.g.][]{too12,too14,kor17,sto22}.

\subsubsection{NS natal kicks} \label{sec:kicks}

It is essential to emphasise that NS formation in a binary system involves two types of kicks: (1) those due to mass loss during the supernova explosion, and (2) an additional NS-specific natal kick. The first type, referred to as `Blaauw' kicks \citep{bla61}, is particularly relevant in close binary systems, where interactions between the stars can strip the donor star's envelope before the supernova event. The `Blaauw' kick magnitude can vary, depending on factors such as mass ratio, initial orbital parameters, and mass loss during the explosion. Generally, its effect is expected to be limited compared to the NS natal kick \citep{hua63,tut73,leo94}. 
The formation mechanism for this additional natal kick remains unsolved \citep[e.g.][]{kus96,sch06,won13,hol17,kat18}, but likely involves anisotropies in neutrino losses and/or mass loss in the supernova ejecta \citep[for a review, see][]{jan12}. Significant progress has been made recently in core-collapse supernova numerical simulations and resulting NS natal kicks \citep[e.g.][]{ColemanBurrows2022MNRAS,BurrowsWang2023arXiv,jan24}.

NS natal kicks were discovered observationally \citep{Lyne1994Natur} and are observed as large peculiar velocities of isolated radio pulsars ($100$~--~$1000$~km~s$^{-1}$), typically at least an order of magnitude higher than the peculiar velocities of NS progenitors. These natal kicks have been investigated using proper motion measurements combined with dispersion measure and electron density models \citep{Arzoumanian2002ApJ,Hobbs2005MNRAS}. Due to advancements in instrumentation -- specifically, the usage of the Very Long Baseline Array with broadband phase modelling \citep[e.g.][]{Brisken2002ApJ} -- it is now possible to measure parallax and proper motions for a large number of isolated radio pulsars \citep{Deller2019ApJ}. An assumption that peculiar velocities of young, isolated radio pulsars represent the natal kicks of NSs has several limitations as discussed by \citet{Igoshev2021MNRAS} and \citet{Mandel2023ApJ}. 
In addition, there is substantial evidence that NSs formed in binaries also receive natal kicks. For instance, the number of X-ray binaries in the Small Magellanic Cloud is much smaller than expected if all NSs received only `Blaauw' kicks (e.g., see \citealt{Igoshev2021MNRAS}). The strength of natal kicks received by NSs formed in binaries is not well-constrained and is a subject of active research \citep{Igoshev2021MNRAS, Willcox2021ApJ, ODoherty2023arXiv}. \citet{Pfahl2002ApJ} argued that NSs in high-mass X-ray binaries with specific eccentricities and orbital periods received natal kicks below 50~km~s$^{-1}$. Observational studies of orbital and spin periods for NSs in Be X-ray binaries suggest two separate populations of NSs with different natal kicks \citep{Knigge2011Natur}.

In binary population synthesis, natal kicks are commonly modelled using the prescription developed by \citet{Fryer2012ApJ}. This prescription relates the remnant mass to the initial stellar mass, and the absolute value of natal kick for core-collapse supernovae is drawn from a single Maxwellian velocity distribution with the distribution parameter $\sigma = 265$~km~s$^{-1}$, as estimated by \cite{Hobbs2005MNRAS}. The natal kicks are drawn uniformly on a sphere. The resulting orbital changes are typically calculated using equations from \cite{Brandt1995MNRAS}. 

Both natal kicks and mass loss kicks contribute to a binary system's systemic velocity following a supernova explosion. We emphasise that the binary's systemic velocity represents the combined motion of both stars within the Galactic reference frame. This velocity accounts for not only the individual motion of each star but also any additional factors such as natal kicks. In the context of NS+WD binary systems, the systemic velocity is influenced by the natal kicks experienced by the NS during its formation and the system's orbital momentum. While some exceptional cases may exhibit systemic velocities of up to 600~km~s$^{-1}$, typical values are expected to be around 150~km~s$^{-1}$ \citep{too18}, also see discussion in Section~\ref{s:syst_velocities}. Our binary population synthesis includes modelling of the systemic velocity.

In this study we examine four alternative supernova kick prescriptions for the NS+WD population models:

\begin{itemize}
    \item `Blaauw' prescription: the kick is calculated as the unbalanced orbital momentum resulting from sudden mass loss \citep{bla61}.
    \item `Hobbs' prescription: the kick is drawn from a Maxwellian distribution \citep{Hobbs2005MNRAS}:
    \begin{equation}
    f_\mathrm{M} (v| \sigma) = \sqrt{\frac{2}{\pi}} \frac{v^2}{\sigma^3} \exp\left[ - \frac{v^2}{2\sigma^2}\right]; \quad  (0< v < \infty)     
    \end{equation}
    with the distribution parameter $\sigma = 265$~km~s$^{-1}$.
    \item `Arzoumanian' prescription: \cite{Arzoumanian2002ApJ} determined the natal kick distribution as the sum of two Maxwellians
    \begin{equation}
    f_\mathrm{2M} (v| w, \sigma_1, \sigma_2) = w f_\mathrm{M} (v| \sigma_1) + (1-w) f_\mathrm{M} (v| \sigma_2) ;   
    \end{equation}
    where $w = 0.4$, $\sigma_1 = 90$~km~s$^{-1}$, and $\sigma_2 = 500$~km~s$^{-1}$.
    \item `Verbunt' prescription: \citet{Verbunt2017AA} determined empirically that the natal kick distribution consists of two Maxwellians with $w = 0.42$, $\sigma_1 = 75$~km~s$^{-1}$, and $\sigma_2 = 316$~km~s$^{-1}$. While this distribution represents the peculiar velocities of young isolated NS well, the physical origin of the low- and high-velocity components is still unclear.  This distribution is our default choice.   
\end{itemize}
Finally, kick directions are assigned isotropically.

\subsubsection{Common envelope evolution} \label{sec:CE}

Observations of double compact objects in compact binaries and their mergers have prompted the concept of a phase that results in substantial orbital shrinkage \citep[e.g.][]{ost73,pac76,van76V,mey79}. This phase may occur when one of the stars within a binary system evolves into a giant, and mass-transfer between the stars may happen if the binary's orbit is comparable in size to the giant star. If the material lost by the donor star (the giant) exceeds the amount the companion star (the accretor) can receive, a runaway process can develop, leading to the companion becoming fully engulfed by the giant's envelope. This unstable, runaway mass-transfer phase is referred to as the {\it common envelope} (CE) phase \citep[for a review, see][]{iva13, rop23}. Conceptually, the orbit of the companion star within the common envelope generates drag -- either gravitational or hydrodynamical. This drag facilitates the transfer of orbital energy and angular momentum to the envelope material, thereby unbinding it from the system. As a result, the companion star spirals inward towards the core of the giant, reducing binary's orbital separation.

In binary population synthesis, the CE phase is often modelled based on energy conservation \citep{pac76,web84,liv88,deK87,deK90}. The binding energy of the envelope, $E_{\rm bind}$, is related to the  loss of orbital energy $\Delta E_{\rm orb}$ through the equation:
\begin{equation}
E_{\rm bind} = \frac{GM_{\rm d}M_{\rm c}}{\lambda R} = \alpha \Delta E_{\rm orb},
\end{equation}
where $M_{\rm d}$ is the donor star's mass, $M_{\rm c}$ is the mass of its core, $R$ is its radius, and $\lambda$ is the structure parameter of its envelope that depends on the structure of the donor. The efficiency of converting orbital energy to unbind the envelope is represented by $\alpha$, which is often chosen based on observational calibrations \citep[e.g.][]{nel00,zor10,sch23}. As a result, the final orbital separations (and, as a consequence, the delay times between the end of the common envelope phase and the end of the gravitational wave inspiral) of the binary populations can vary depending on this choice. A higher $\alpha$ value corresponds to a more efficient envelope ejection. 

The alternative model for CE evolution, known as the $\gamma$-CE, focuses on the balance of angular momentum rather than that of energy \citep{nel00}. It is represented as follows:
\begin{equation}
    \frac{\Delta J}{J} = \gamma \frac{\Delta M}{M},
\end{equation}
where $\Delta J$ and $\Delta M$ are respectively the angular momentum and mass loss. The $\gamma$-prescription was introduced to account for the first phase of mass-transfer during the formation of WD+WD binaries with $\gamma = 1.75$ \citep{nel00,nel05,vdS06}.

Based on these two common envelope prescriptions, we construct two families of NS+WD population models: $\alpha\alpha$ and $\gamma\alpha$. In model $\alpha\alpha$, the $\alpha$-formalism is used to determine the outcome of all CE phases. In $\gamma\alpha$ models, the $\gamma$-prescription is applied unless the binary contains a compact object or if the CE is triggered by a tidal instability (rather than dynamically unstable Roche lobe overflow, see \citealt{too12}). Thus, in the $\gamma \alpha$ models the first CE phase is typically modelled with the $\gamma$-CE prescription; the second CE (with a giant donor and white dwarf or neutron star companion) is typically described by the $\alpha$-formalism. Our default model assumes $\alpha\lambda = 2$ (referred to as $\alpha\alpha$) and also considers a variation with $\alpha\lambda = 0.25$ (referred to as $\alpha\alpha2$). Our $\alpha\alpha$ model, calibrated on observed WD+WDs \citep[specifically the second mass-transfer phase, see][]{nel00,nel01}, contrasts with our $\alpha\alpha2$ model, which is calibrated on the formation of compact WDs in binaries with M type main-sequence stars \citep[][]{zor10,too13,cam14,zor14}. 
We recognise that both CE prescriptions have their own uncertainties and limitations, and they are often chosen based on the specific context of the binary system being modelled. Here we employ both to better understand their implications on the NS+WD population detectable with {\it LISA}.

\subsection{From NS+WD formation to present time} \label{sec:MSMStopresent}

The following stage in our modelling involves adjusting the properties of the simulated NS+WD binaries (such as their spatial, orbital separation, and eccentricity distributions) to resemble those of the present-day Milky Way's stellar population. This process necessitates two distinct types of integration, each carried out independently as detailed below. The first integration evolves binary orbital parameters (semi-major axis and eccentricity), which gradually decay due to GW radiation, thereby circularising the binary. The second integration involves the NS+WD orbit within the Galactic potential. The latter depends on the choice of the Milky Way's gravitational potential, while the former relies on the selection of its star formation history.

\subsubsection{NS+WD present-day orbital parameters} \label{sec:orb_evol}

In this study, we adopt a star formation history from the chemo-spectrophotometric model of the Milky Way by \citet[][see also \citealt{BP00}]{BP99}. This model employs an `inside-out' formation scheme for the disk and incorporates empirically and/or theoretically justified prescriptions for the star formation rate, including metallicity-dependent stellar properties. We note, however, that our binary population models assume solar metallicity, which simplifies the complex chemical enrichment history of the Milky Way. This choice is somewhat justified by the observation that, when modelling \textit{LISA}-detectable Galactic WD+WD population, variations due to different CE assumptions are significantly larger than those resulting from variations in other factors, such as the initial mass function, metallicity, and binary fraction \citep{kor20}. We distribute NS+WD binaries in time (and space, see Section \ref{sec:MWmodel}) according to their (spatially resolved) star formation grid, which determines the binary's age. As in \citet{BP99}, we assume the Milky Way's age to be 13.5\,Gyr.

Subsequently, we determine the present-day orbital parameters of NS+WDs—semi-major axis $a$ and eccentricity $e$—by accounting for GW radiation reaction from NS+WD formation until 13.5\,Gyr. To do so, we numerically solve equations originally derived in \citet{pet64}:
\begin{equation}
a(e) = c_0 \frac{e^{12/19}}{(1-e^2)} \left( 1 + \frac{121}{304} e^2 \right)^{870/2299},
\end{equation}
and
\begin{equation}
\frac{de}{dt} = -\frac{19}{12} \frac{\beta}{c_0^4} \frac{e^{-29/19}(1-e^2)^{3/2}}{ \left[1+ (121/304)e^2 \right]^{1181/2299}},
\end{equation}
where $c_0$ is determined by the initial condition $a(e_0)=a_0$, with $a_0$ and $e_0$ representing the binary's semi-major axis and eccentricity at NS+WD formation, and
\begin{equation}
\beta = \frac{64}{5}\frac{G^3 m_1 m_2 (m_1+m_2)}{c^5}
\end{equation}
with $c$ being the speed of light. Lastly, we exclude binaries if their formation times exceed 13.5 Gyr, if they have initiated mass-transfer (i.e., when the WD fills its Roche lobe), or if they have already merged within this time frame. We deffer the modelling of interacting NS+WD binaries to a future study.

\subsubsection{NS+WD present-day spatial distribution} \label{sec:MWmodel}

Our goal is to investigate the size and distribution of properties of NS+WDs detectable with {\it LISA}, so we can disregard the detailed structure of our Galaxy, such as the shape of its central bulge/bar region, the structure of the stellar disk including spiral arms, and the stellar halo component. We assume an analytical expression for the gravitational potential consisting of three main components: bulge, stellar disk, and dark matter halo.

To model the star formation history of the disk, as in \citet{nel04}, we linearly interpolate the plane-projected star formation rate of \citet[][SFR$_{\rm BP}(R,t)$]{BP99}, which extend up to 13.5\,Gyr in time and up to 19\,kpc in radius. We assume that the probability of a binary being born at a radius smaller than $R$ follows the integrated SFR, defined as:
\begin{equation} \label{eqn:SFR_BP}
    P(R,t) = \frac{\int_0^R {\rm SFR}_{\rm BP}(R',t) 2\pi R' dr'}{\int_0^{19} \rm{SFR}_{\rm BP}(R',t) 2\pi R' dr'},
\end{equation}
where $0 \le R \le 19$\,kpc is the cylindrical radius measured from the Galactic Centre. Additionally, we assume the vertical distribution of stars that follows a $\sech^2$ function. As a result of the assumptions above, our disk stellar number density profile can be analytically described as
\begin{equation}  \label{eqn:disk}
\rho_{\rm disk}(R,z) \propto e^{-R/R_{\rm d}} \sech^2 \left( \frac{z}{z_{\rm d}} \right) \ {\text{kpc}}^{-3},
\end{equation} 
where $R_{\rm d} = 2.5\,$kpc and $z_{\rm d}=300$\,pc are respectively the characteristic scale radius and scale height of the disk \citep[][]{BP99}. By integrating SFR$_{\rm BP}(R,t)$ up to 13.5\,Gyr, we obtain a total mass of $5 \times 10^{10}\,$M${}_{\odot}$, which is consistent with recent studies \citep[e.g.][]{lic15}. We note that the disk scale parameters (total stellar mass, scale radius and scale heights) in our model fall within the range of values derived from observations \citep[see][for a review]{bla16}. We assume the distance of the Sun from the Galactic Centre to be $R_{\odot}=8.5\,$kpc and $V_\odot = 240$~km~s$^{-1}$ compatible with \cite{Feast1997MNRAS} and \cite{Reid2014ApJ}. The results by \cite{GRAVITY19} are slightly different $R_{\odot} = 8.178\pm 0.013_\mathrm{stat} \pm 0.022_\mathrm{syst}$~kpc. However, we anticipate that this  difference does not affect our results. In our experience this difference could result in $\approx 10$~km~s$^{-1}$ difference which is very small in comparison to typical systematic velocities of binaries.

We model the bulge component by doubling the star formation rate in the inner $3\,$kpc of the Galaxy and distributing binaries as
\begin{equation} \label{eqn:bulge}
\rho_{\rm bulge}(r) \propto e^{-r^2/2 r_{\rm b}^2}\ {\text{kpc}}^{-3},
\end{equation}
where $r$ is the spherical distance from the Galactic Centre and $r_{\rm b}=0.5\,$kpc is the characteristic radius. Although our modeling approach is somewhat simplistic, it is based on the inside-out star formation process described by \citet{BP99}. Consequently, in our representation of the Milky Way, we find that the median age of binaries is approximately 10 Gyr in the bulge, decreasing to around 3 Gyr at the Solar radius, and decreasing further down to about 2 Gyr at the disk's outskirts, as expected from observations \citep[e.g.][]{hay16,mak17,gra20}. We find the integrated total mass of the bulge ($r<3$ kpc) at the present time is $\sim2.6 \times 10^{10}\,$M${}_{\odot}$.

To model the density distribution of the dark matter halo we use the Navarro-Frenk-White profile \citep{NFWprofile}:
\begin{equation} \label{eqn:halo}
\rho_{\rm DM} (r) = \frac{\rho_{\rm h}}{(r/r_{\rm s})(1+r/r_{\rm s})^2} \ {\text M}_{\odot}\, {\text{kpc}}^{-3},
\end{equation}
where $r_{\rm s} = 20\,$kpc is the scale length of the halo and $\rho_{\rm h} = 0.5 \times 10^7\,$M$_{\odot}$~kpc$^{-3}$ is the halo scale density \citep[e.g.][]{nes13}.
The total mass of the halo can be obtained by integrating Eq.~\eqref{eqn:halo} from the centre to the maximum Galactocentric radius  of 100$\,$kpc, which for our fiducial parameters yields $4.8 \times 10^{11}\,$M$_{\odot}$.

\begin{figure}
	\includegraphics[width=1.1\columnwidth]{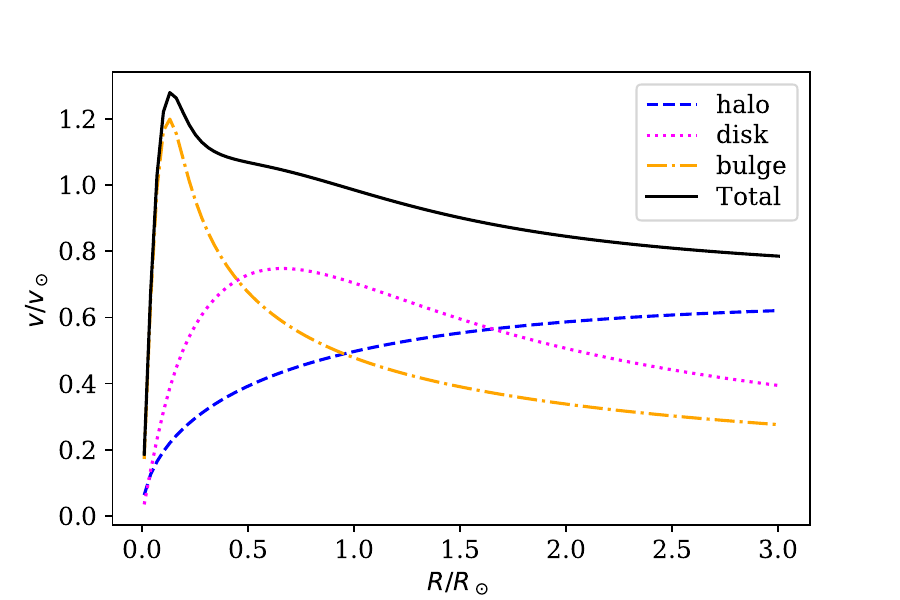}
    \caption{The total rotation curve of the Milky Way is represented by the solid black line, with the contributions from individual components depicted as follows: disk (magenta dotted line), bulge (gold dash-dotted line), and dark matter halo (blue dashed line). In our simulations, the Sun's distance from the Galactic Centre, $R_\odot$, and its circular velocity, $V_\odot$, are set at 8.5\,kpc and 240\,km s$^{-1}$, respectively.} 
   \label{fig:rotationalProfile}
\end{figure}

We construct a static gravitational potential based on the present-day Galactic properties following the same model as described in Eqs. (\ref{eqn:disk}-\ref{eqn:halo}) with the help of {\sc galpy} package \citep{Bovy2015ApJS}. 
We compute the amplitudes of individual components of the {\sc galpy} potential as the following:
\begin{equation}
A_\mathrm{bulge} = \frac{M_{\rm b}}{\left(\sqrt{2\pi} r_{\rm b}\right)^3},    
\end{equation}
\begin{equation}
A_\mathrm{disk} = \frac{M_\mathrm{total}}{4\pi h_{\rm z} h_{\rm r}^2},    
\end{equation}
\begin{equation}
A_\mathrm{DM} = 4\pi r_{\rm s}^3 \rho_{\rm h}.    
\end{equation}
These amplitudes have different dimensionality depending on the structure of the respective component and how exactly the aforementioned mathematical expressions are implemented in {\sc galpy}\footnote{ See the documentation for more details \url{https://docs.galpy.org/}}. These equations are summarised here to make our work reproducible. We show the resulting rotational curve and contribution of individual Galactic components to the total gravitational potential in Figure~\ref{fig:rotationalProfile}.  

We prepare the initial conditions for the Galactic orbits integration, from receiving the NS kick until the present time, as follows. We remind the reader that binary's birth time is determined by Eq.~\eqref{eqn:SFR_BP}, while the time necessary until NS formation is provided as part of our binary population synthesis modelling. Binary's coordinates are sampled randomly following the Milky Way density profile (cf. Eqs.~\ref{eqn:disk}-\ref{eqn:halo}). The initial velocities are the sum of three components: (1) the speed of the local standard of rest (LSR), $v_\mathrm{LSR}$, (2) a small peculiar velocity, $v_\mathrm{p}$, of the progenitor binary, and (3) a randomly oriented kick, $v_\mathrm{k}$, received by the binary due to the NS natal kick and mass loss.

The components of the binary peculiar velocity are drawn from a normal distribution, $v_\mathrm{p} \sim \mathcal{N}(0, \sigma)$, with $\sigma = 10$~km/s in each direction, see e.g. \cite{RamrezTannus2021AA} for typical velocity dispersion of massive stars. To model the components of the binary kick, we draw two angles, $\theta \in [0, 2\pi]$ and $\phi \in [0, \pi]$, describing the orientation of the kick. These angles are drawn uniformly on a sphere i.e. we ensure the isotropic velocity distribution. The individual kick components are then computed as:
\begin{equation}
\begin{array}{ccl}
v_\mathrm{k,r}  &=& v_k \sin \phi \cos \theta;  \\
v_\mathrm{k,t}  &=& v_k \sin \phi \sin \theta; \\
v_\mathrm{k,z}  &=& v_k \cos \phi.
\end{array}    
\end{equation}
Overall, the components of the total velocity ($v_r$ is radial component along the axis connecting Galactic Centre and star's position and pointing away from the Galactic Centre, $v_t$ is component along the Galactic rotation and $v_z$ is a component along the direction perpendicular to the disk and aligned with the Galactic spin axis) can be described as the following:
\begin{equation}
 \begin{array}{ccl}
v_r &=& v_\mathrm{k,r} + v_\mathrm{p,r};  \\
v_t &=& v_\mathrm{k,t} + v_\mathrm{p,t} + v_\mathrm{LSR}; \\     
v_z &=& v_\mathrm{k,z} + v_\mathrm{p,z}. \\
 \end{array}   
\end{equation}
We integrate the orbit of a binary in the Galactic gravitational potential from NS formation 
until present using the Dormand-Prince integrator \citep{Dormand1980} as implemented in {\sc galpy} using the components of Galactic potential described above. Here we assume that the gravitational potential does not evolve with time. The lengths of the integration interval is computed as 13.5\,Gyr minus the time at which NS formed that we determine as the difference between the time at which the binary has been seeded in the Milky Way (cf. Section~\ref{sec:orb_evol}) and the time required to form the NS binary component (provided by {\sc SeBa}, cf. Section~\ref{sec:MStoNS+WD}). At the end of the integration we check if the total energy is conserved at a level better than $10^{-3}$ of the initial energy. We do not expect that the time-dependent components of the Milky Way such as the bar and the spiral arms can significantly affect the integrated orbits because most of our binaries receive a significant systemic velocity at the moment of supernova explosion and thus they are quite separated in the phase-space from other Galactic components and thus should interact less with them. Nevertheless, these effects should be studied in more details in future publications, especially for bulge population. We also ignore the influence of the Large Magellanic Cloud -- the most massive of the Milky Way's satellites -- on the Galactic gravitational potential \citep[for example, see][]{1avn67, con21}, which could possibly affect binaries' orbits in the Galactic halo. In rare cases where energy is not conserved at the required level, we disregard the orbit. These cases typically correspond to numerically challenging orbits, which may, for example, cross the Galactic Centre. The total number of disregarded orbits is usually below approximately 1\,per cent of the total. We store these present-day positions of binaries in both Cartesian and Ecliptic coordinate systems and assess their detectability with {\it LISA} as detailed below.

\section{Detectability with LISA} \label{sec:lisa_snr_PE_sec}

The majority of double compact object binaries within the {\it LISA} band are expected to have circular (binary) orbits. This is because of two factors. Firstly, to reach short-period orbits binaries undergo from a few to several mass-transfer phases, at least one of these phases is the CE. It is generally assumed that the CE evolution is very efficient in circularising binary's orbit. Secondly, although at a slower rate, GWs radiation also contributes to the circularisation process. However, in cases where the binary system forms in close proximity to, or even directly within, the \textit{LISA} band, the binary can retain some eccentricity if the WD component forms before the NS (cf. Section~\ref{sec:channels}). This residual eccentricity can be attributed to the natal kick the NS receives at the moment of the supernova explosion (cf. Section~\ref{sec:kicks}).

In the context of the {\it LISA} mission, to describe the GW radiation emitted by a typical quasi-monochromatic circular binary eight parameters are needed. These typically chosen to be:
\begin{equation}
    \{\mathcal{A}_0, f_{\rm GW}, \dot{f}_{\rm GW}, \lambda, \beta, \iota, \psi, \phi_0\}\,, 
    \label{eq:params} \nonumber
\end{equation}
where $\mathcal{A}_0$ is the GW amplitude, $(\lambda, \beta)$ are the ecliptic longitude and latitude, respectively, $\iota$ is the inclination angle, $\psi$ is the polarisation angle, and $\phi_0$ is an initial phase. In the circular case, the GW frequency is twice binary orbital frequency $f_{\rm orb}$
\begin{equation}
    f_{\rm GW}=2f_{\rm orb},
\end{equation}
while the GW amplitude is given by
\begin{equation}
    \mathcal{A}_0 = \frac{2 (G \mathcal{M})^{5/3} }{c^4 d} (\pi f_{\rm GW})^{2/3}.
\end{equation}
This is set by the source's distance $d$ and chirp mass
\begin{equation}
    \mathcal{M} = \frac{(m_1 m_2)^{3/5}}{(m_1 + m_2)^{1/5}}, 
\end{equation}
for component masses $m_1$ and $m_2$. The chirp mass also sets the rate at which the frequency changes due to the gravitational radiation reaction:
\begin{equation}
\label{fdot}
    \dot{f}_{\rm GW} = \frac{96}{5}\frac{(G\mathcal{M})^{5/3}}{\pi c^5} (\pi f_{\rm GW})^{11/3}.
\end{equation}

In the eccentric case, binaries emit GWs at multiple frequency harmonics; for a Keplerian orbit these are given by
\begin{equation}
    f_n = \left(\frac{n}{2}\right)f_{\rm GW}
\end{equation}
(with small corrections if the orbit is precessing).
Each harmonic has an amplitude 
\begin{equation}
{\cal A}_n = {\cal A}_0\left(\frac{2}{n}\right)^{5/3}g(n,e)^{1/2},  
\end{equation}
where the function $g(n,e)$ is defined in \citet{pet64}. Importantly, it should be noted that in the case of a circular binary ($e=0$), the function $g(2,0) = 1$, as expected. The rate of frequency change becomes
\begin{equation}\label{eq:fdot}
\dot{f}_{n}=\frac{96}{5}\frac{(G\mathcal{M})^{5/3}}{\pi c^5} (\pi f_{\rm n})^{11/3} F(e),
\end{equation}
where $F(e)$ is the enhancement factor \citep{pet64}
\begin{equation} \label{eqn:fe}
F(e) = \frac{1+\frac{73}{24}e^2+\frac{37}{96}e^4}{(1-e^2)^{7/2}}.
\end{equation}

\subsection{Extracting detectable binaries from the Galactic foreground} \label{sec:LISApipeline}

To get an estimate of the individually detectable NS+WD binaries from the mock catalogue, we use the analysis pipeline presented in \citet{kar21}. It is based on an signal-to-noise (SNR, $\rho$) evaluation using an iterative scheme for the estimate of the confusion foreground generated by Milky Way's WD+WD and NS+WD populations \citep[see also][]{tim06,cro07,nis12}.

This scheme begins with the generation of the signal measured by {\it LISA}, by computing the waveforms for each of the simulated catalogue entries, and by projecting it on the {\it LISA} arms for a given duration of the mission. After the data generation part, we begin with an iterative source subtraction process, each iteration initialised by computing an estimate of the Power Spectral Density (PSD) $S_{ \rm n,\,k}$, for the total noise of the instrument. The index $k$ represents the algorithm iteration number. The $S_{ \rm n,\,k}$ represents the overall noise, which includes the combined effect of overlapping unresolved GW sources and the instrumental noise, and is given by computing the running median of the data PSD. Then we process by computing the SNR $\rho_i$ of each source $i$ using the smoothed $S_{\rm n,\,k}$. The SNR $\rho$ is computed as 
\begin{equation}
    \rho^2 =  \left( h | h \right),
\label{eq:snr} 
\end{equation}
where $h$ is the true waveform template of a circular quasi-monochromatic source, and the $\left( \cdot | \cdot \right)$ denotes the noise weighted inner product in frequency domain, which for two real time series $a$ and $b$ is written as:
\begin{equation}
    \left( a | b \right) = 2 \int\limits_0^\infty \mathrm{d}f \left[ \tilde{a}^\ast(f) \tilde{b}(f) + \tilde{a}(f) \tilde{b}^\ast(f) \right]/S_n(f).
\label{eq:ineerprod} 
\end{equation}
The `$\,\tilde{\,\,}\,$' here represents the Fourier transformation, and the `$\,^\ast$\,' the complex conjugation. The $S_n$ is the one-sided noise PSD.

If $\rho_i > \rho_0$, where $\rho_0$ a given chosen SNR threshold, the source is classified as resolvable, and is thus subtracted from the data. The smoothed PSD of the residual $S_{\rm n,\,k+1}$ is re-evaluated after iterating over the catalogue of sources, and the procedure is repeated until the algorithm converges. Convergence is reached when all sources are subtracted given the $\rho_0$ threshold, or if $S_{\rm n,\,k+1}$ and $S_{\rm n,\,k}$ are practically identical at all frequencies considered. At the end of this procedure, we compute the final SNR with respect to the final estimate of $S_{ \rm n,\,k_\mathrm{final}}$, for the sources recovered.

For every binary listed in our data sets, we conduct an analysis using the Fisher information matrix (FIM) to gauge the precision of parameter extraction. It is important to note that the error assessments from the FIM analysis hold true predominantly for large SNR values \citep[\( \rho \gtrsim 20 \), for instance][]{cut98}. Therefore, in some cases, the derived uncertainties may be underestimated. A comprehensive Bayesian parameter evaluation is essential to ascertain more realistic uncertainties for the binary parameters \citep[e.g.][]{katz22}.

Finally, we should mention again that {\it LISA} is going to measure thousands of signals originating from different types of GW sources (Galactic, extra-galactic and cosmological), overlapping in time and in frequency. This demands for taking into account the correlation between the different waveforms and their corresponding parameters. Thus, a {\it Global Fit} pipeline, based on costly computational algorithms~\citep[e.g.][]{lit23}, is necessary. The procedure described in this section can be considered as a simulation of a computationally costly {\it Global Fit} analysis of the {\it LISA} data, and is necessary in order to keep the computational cost in acceptable levels at this exploratory stage.

\subsection{Eccentricity measurement} \label{sec:ecc}

Depending on the binary parameters, not all harmonics may reach the detectability SNR threshold, which is described in the following section.
If no harmonics reach this threshold then the source is not detectable. 
If just one harmonic reaches this threshold then the source is detectable, but might be mistaken for a source on a circular orbit (e.g.\ a WD+WD binary).
If two (or more) harmonics reach the threshold then they can both (all) be detected individually and, when analysed in combination, used to measure the eccentricity. 
Therefore, requiring that a source has at least two harmonics above the SNR threshold gives a simple, conservative criteria for classifying a binary as having a detectable (i.e.\ nonzero) eccentricity.

The above criterion for identifying an eccentric source is conservative in the sense that it might still be possible to measure a nonzero eccentricity even when the second (loudest) harmonic is slightly below the threshold SNR. The possibility was investigated in detail by \cite{Mooreetal} who were able to classify sources with detectabile eccentricities based on the frequency of the dominant ($n=2$) harmonic, $f_{\rm GW}$, and the total SNR, summed across all harmonics, $\rho$. Considering NS+WD binaries, the minimum detectable eccentricity was found to be \citep{Mooreetal}
\begin{align} \label{eq:analytic_emin}
    e_\mathrm{min}(f_\mathrm{GW}, \rho) \approx &\left(\frac{1}{\rho^{1.54}} + \frac{1}{\rho} \right) \nonumber \\
    \times \{1.08 &+ 0.87\tan^{-1}\left[1.08 \left(f_\mathrm{GW}/\mathrm{mHz} - 2.13 \right)\right] \nonumber \\
    &- 0.55\tan^{-1}\left[2.08 \left(f_\mathrm{GW}/\mathrm{mHz} - 1.22 \right)\right]\}. 
\end{align}
This fit has been tested for $0.5 \leq f_{\rm GW}/\mathrm{mHz} \leq 10$ and $\rho \geq 8$. These results was derived using the \textsc{Balrog} software package that are being developed for {\it LISA} data analysis and parameter estimation for all source types: supermassive binary black hole mergers \cite{2023PhRvD.107l3026P}, WD+WD binaries \citep{bus19,2020ApJ...894L..15R,fin22}, and stellar-mass binary black holes \citep{2021PhRvD.104d4065B,2022arXiv220403423K, 2023arXiv230518048B}.


\section{Results}

\begin{figure}
	\includegraphics[width=\columnwidth]{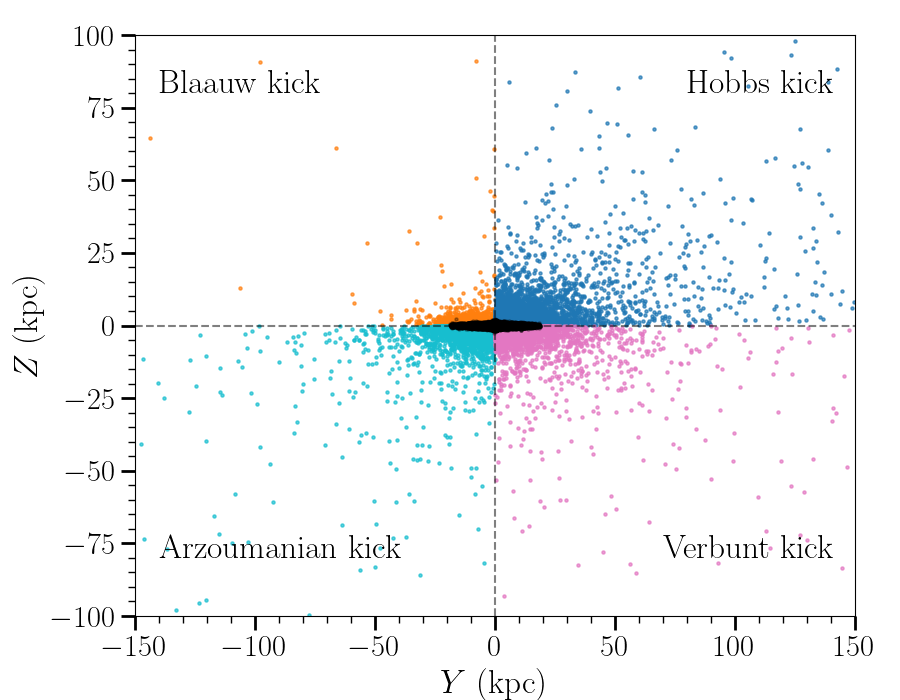}
    \caption{Comparison of present-day NS+WD positions in $Y-Z$ Galactocentric coordinates under varying NS natal kick prescriptions, depicted in different colours. The initial (MS+MS) positions are indicated in black for reference.} 
   \label{fig:kicks}
\end{figure}
\begin{figure*}
	\includegraphics[width=1.8\columnwidth]{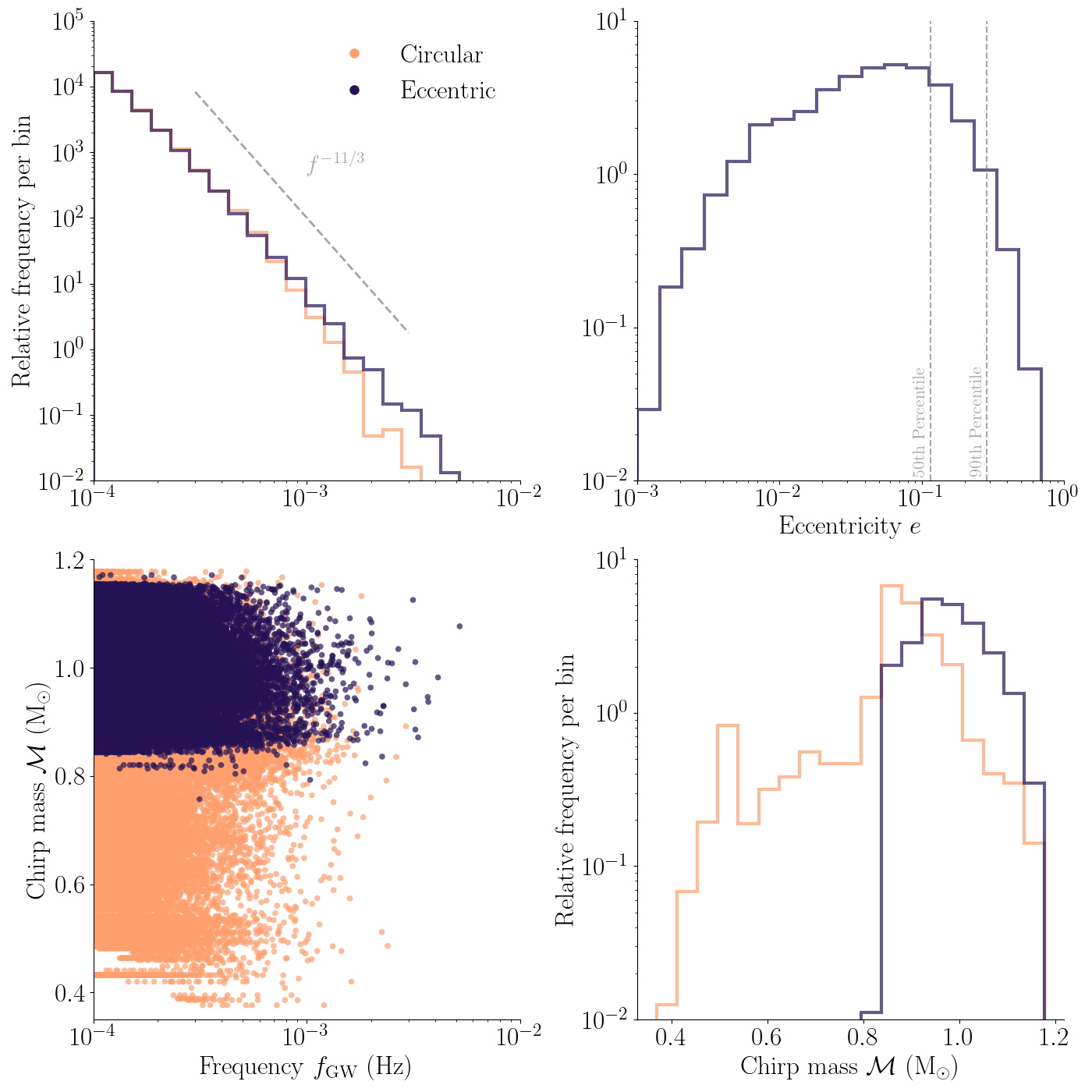}
    \caption{Displayed is our fiducial model of the present-day Galactic NS+WD population ($\alpha\alpha$-CE + `Verbunt' kick), represented in the frequency - chirp mass parameter space (bottom left), with projected frequency and chirp mass distributions per bin presented in the top left and bottom right panels respectively. Binaries are categorised by colour: circular ($e=0$, pale orange) and eccentric ($e\neq0$, purple). The top right panel showcases the distribution of eccentricities. The dashed grey line in the top left panel shows the expected $\propto f^{-11/3}$ distribution in frequency, with fewer sources at higher frequencies because of the accelerating, or chirping, inspiral described in Eq.~\eqref{eq:fdot}.}  \label{fig:mfe}
\end{figure*}
\subsection{Overall NS+WD population in the mHz band}

Based on our modelling results (cf. Section~\ref{sec:methods}) we estimate that there are between $\mathcal{O}(10^4)$ and $\mathcal{O}(10^5)$ NS+WD binaries currently emitting GWs in the {\it LISA} frequency band of $(10^{-4}-10^{-1})$\,Hz within the Milky Way. The exact number is dependent on the assumptions made for the NS natal kick and CE prescriptions (see Table~\ref{tab:numbers}), as discussed in the following. We anticipate that these differences becomes more pronounced after we apply {\it LISA}'s selection effects (cf. Section~\ref{sec:LISAdetetctable}). Our fiducial simulation using the $\alpha\alpha$-CE prescription with $\alpha\lambda = 2$ and the `Verbunt' NS natal kick prescription generates $1.75\times10^5$ mHz NS+WD binaries. We highlight that this is two orders of magnitudes lower compared to the total number of WD+WD binaries in the {\it LISA} band, which is also reported in Table~\ref{tab:numbers} for comparison \citep[see also][]{LISAastro}.
We observe a significant decrease in the number of mHz NS+WD binaries for $\alpha\alpha2$ and $\gamma\alpha$ CE models \citep[see also][]{chu06,too18}. The number of NS+WDs decreases by almost 90\,per~cent for the $\alpha\alpha2$-CE model, and by 30\,per cent for the $\gamma\alpha$-CE model. 

Table~\ref{tab:numbers} also indicates that the assumption regarding the NS natal kick significantly influences our results. Specifically, when comparing models using the same CE prescription and efficiency, we find that variations with the `Blaauw' kick prescription result in the highest number of NS+WDs within the LISA band. This outcome is expected, as in this case, the kick velocities are the lowest compared to other prescriptions. Even modest natal kicks can provide enough energy for the binary to move to large distances  beyond the Galactic disk or, in extreme scenarios, to completely unbind the binary from the Galaxy. Figure~\ref{fig:kicks} showcases the impact of various natal kick prescriptions on the spatial distribution of NS+WD in the Milky Way. We emphasise the positions of the binaries at formation in black, and their present-day positions in colour. It is clear that even with the application of the `Blaauw' prescription (top left), NS+WD binaries are significantly dispersed at larger distances compared to their initial spatial distribution. This effect is most pronounced for the `Hobbs' kick prescription. We examine this in more detail in Section~\ref{sec:Zdist}.


\begin{table*}
\caption{Number of WD+WD and NS+WD binaries in the Milky Way based on various model assumptions. In particular, we consider results presented in Column A displays the total number of NS+WD in the {\it LISA} band ($10^{-4} - 10^{-1}\,$Hz); Column B indicates count of detectable eccentric ($e\neq0$) NS+WDs; columns C and D specify the number of NS+WDs with measurable eccentricity using two different methods (cf. Section~\ref{sec:ecc}). Specifically, column C presents results from detecting at least two harmonics of the binary with $\rho_{\rm 4yr}>7$, whereas column D uses the result from applying Eq.~\eqref{eq:analytic_emin}.
}\label{tab:numbers}
\begin{tabular}{lll|cc|ccc|cl}
\hline
\multirow{3}{*}{CE}              & \multirow{3}{*}{CE efficiency}               & \multirow{3}{*}{NS kick} & \multicolumn{2}{c|}{In the {\it LISA} band}                                                                                                           & \multicolumn{3}{c|}{Detectable}                                                                                                                                                                                        & \multicolumn{2}{c}{Measured $e$}                                            \\ \cline{4-10} 
                                 &                                       &                          &                                                                                  & A                                                            &                                                                                  &                                                      & B                                                                            & \multicolumn{2}{c}{(C, D)}                                                   \\
                                 &                                       &                          & \multicolumn{1}{l}{\begin{tabular}[c]{@{}c@{}}WD+WD\\ $ \times10^7$\end{tabular}} & \begin{tabular}[c]{@{}c@{}}NS+WD\\$\times10^5$\end{tabular} & \multicolumn{1}{c}{\begin{tabular}[c]{@{}c@{}}WD+WD\\ 
 $\times10^3$\end{tabular}} & \begin{tabular}[c]{@{}c@{}}NS+WD\\ $e=0$\end{tabular} & \multicolumn{1}{c|}{\begin{tabular}[c]{@{}c@{}}NS+WD\\ $e\neq0$\end{tabular}} & \multicolumn{2}{c}{\begin{tabular}[c]{@{}c@{}}NS+WD\\ $e\neq0$\end{tabular}} \\ \hline
\multirow{4}{*}{$\alpha\alpha$}  & \multirow{4}{*}{$\alpha\lambda=2.00$} & Verbunt                  & \multirow{4}{*}{2.08}                                                                           & 1.75                                                         & 22.95                                                                            & 62                                                   & 43                                                                           & \multicolumn{2}{c}{(5, 25)}                                                 \\
                                 &                                       & Arzoumanian              &                                                                                  & 1.32                                                         & 22.97                                                                            & 64                                                   & 44                                                                           & \multicolumn{2}{c}{(6, 26)}                                                \\
                                 &                                       & Hobbs                    &                                                                                  & 0.76                                                         & 22.96                                                                            & 31                                                   & 44                                                                           & \multicolumn{2}{c}{(6, 17)}                                                \\
                                 &                                       & Blaauw                   &                                                                                  & 6.39                                                         & 22.94                                                                            & 246                                                  & 22                                                                           & \multicolumn{2}{c}{(3, 6)}                                                 \\ \hline
\multirow{4}{*}{$\alpha\alpha2$} & \multirow{4}{*}{$\alpha\lambda=0.25$} & Verbunt                  & \multirow{4}{*}{2.40}                                                            & 0.23                                                         & 0.75                                                                             & 22                                                   & 128                                                                          & \multicolumn{2}{c}{(27, 45)}                                               \\
                                 &                                       & Arzoumanian              &                                                                                  & 0.18                                                         & 0.75                                                                             & 10                                                   & 123                                                                          & \multicolumn{2}{c}{(22, 49)}                                                \\
                                 &                                       & Hobbs                    &                                                                                  & 0.14                                                         & 0.75                                                                             & 8                                                    & 107                                                                          & \multicolumn{2}{c}{(37, 58)}                                                \\
                                 &                                       & Blaaw                    &                                                                                  & 0.53                                                         & 0.75                                                                             & 140                                                  & 217                                                                          & \multicolumn{2}{c}{(28, 68)}                                               \\ \hline
$\gamma\alpha$                   & $\alpha\lambda=2.00$, $\gamma=1.75$   & Verbunt                  & 2.87                                                                             & 1.24                                                         & 24.45                                                                            & 35                                                   & 47                                                                           & \multicolumn{2}{c}{(9, 29)}                                                \\ \hline
\end{tabular}
\end{table*}

\begin{figure*}
 	\includegraphics[width=1.8\columnwidth]{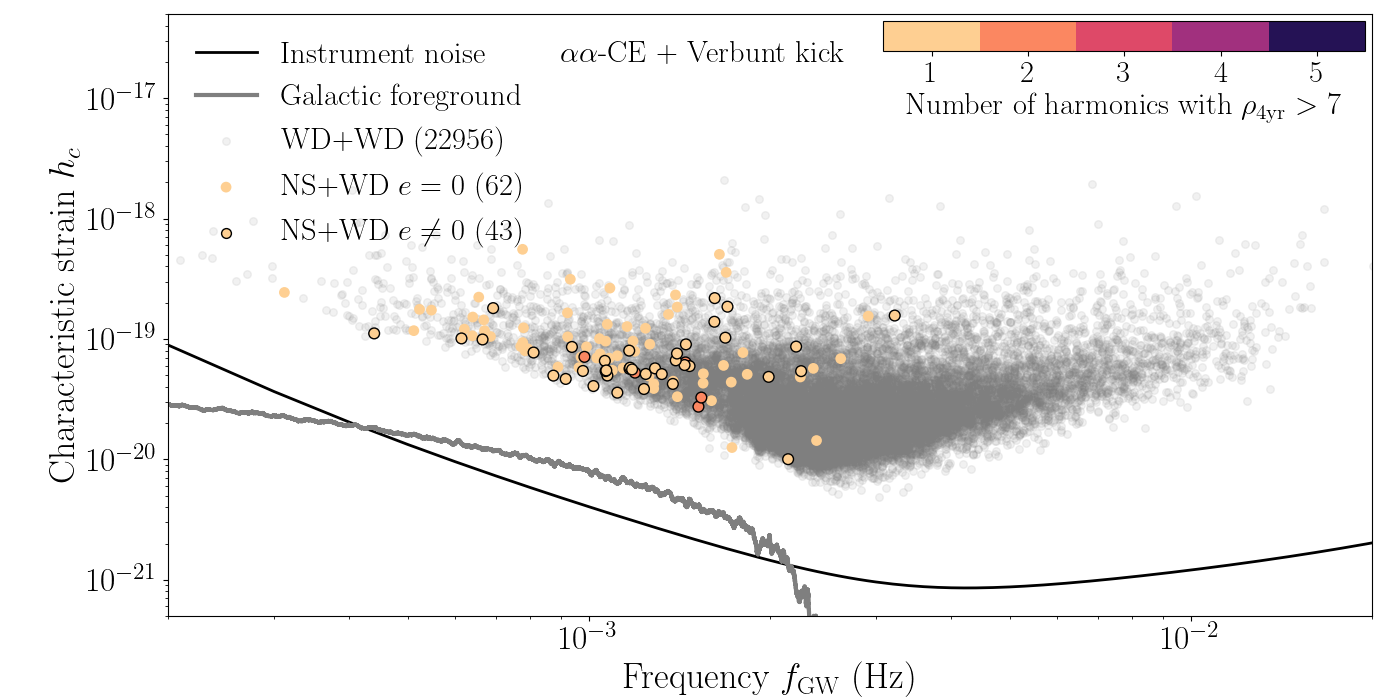}
   	\includegraphics[width=1.8\columnwidth]{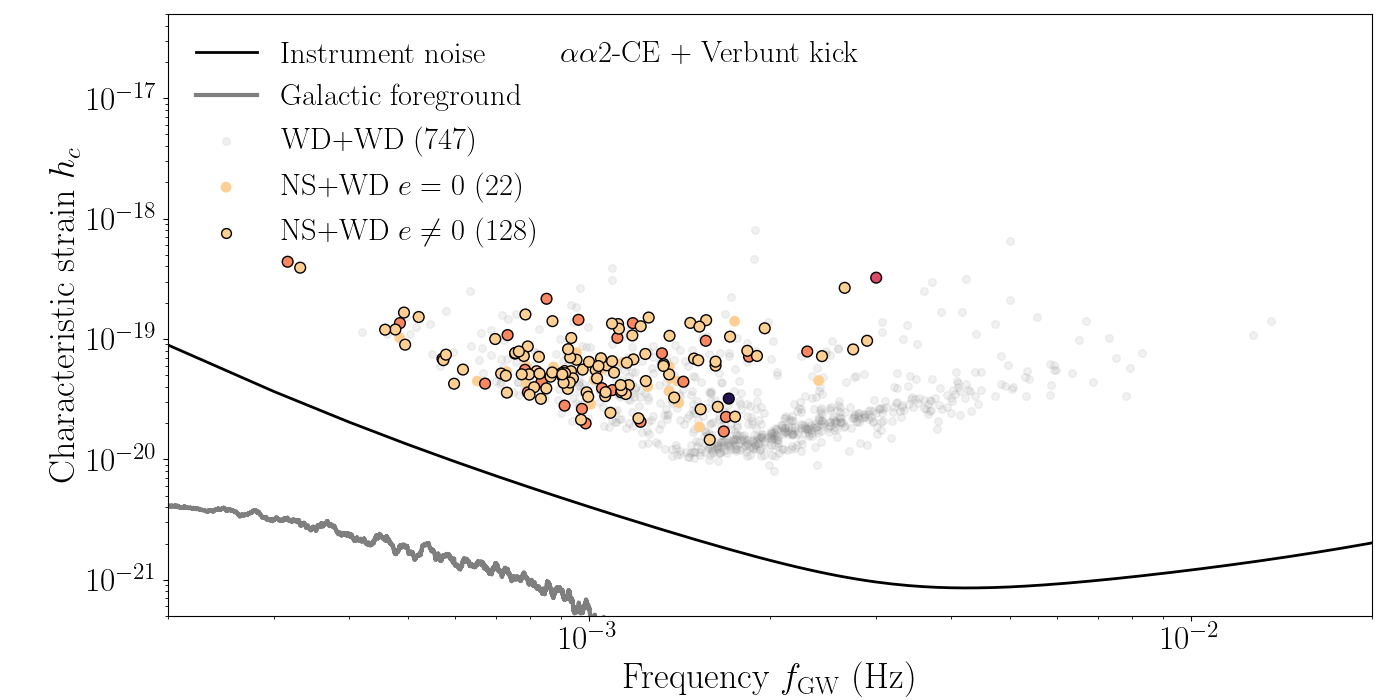}
    \caption{{\it LISA}-detectable Galactic NS+WD (in colour) and WD+WD (in grey) populations in the GW frequency-characteristic strain parameter space. The top panel presents our fiducial model ($\alpha\alpha$-CE with $\alpha\lambda = 2$ and the `Verbunt' natal kick), while the  bottom panel displays an alternative model ($\alpha\alpha2$-CE with $\alpha\lambda = 0.25$ and the `Verbunt' natal kick). We only plot binaries that have SNRs of $\rho_{\rm 4yr}>7$. NS+WD binaries with $e\neq0$ are highlighted with a black contour. For these eccentric binaries, individual harmonics must meet the SNR criterion. For clarity, only the highest SNR harmonic for each binary is shown, with the colour indicating the number of detectable harmonics specific to that binary. The black solid line showcases {\it LISA}'s sensitivity curve \citep{LISAscireq}. The grey solid line illustrates the unresolved Galactic foreground, calculated self-consistently from the two populations (NS+WD and WD+WD binaries) for a 4-year mission duration. 
    }  \label{fig:hcf}

\end{figure*}
\subsubsection{Main formation pathways} \label{sec:channels}

Figure~\ref{fig:mfe} displays the present-day Galactic NS+WD population in the frequency--chirp mass parameter space. We can visually identify two distinct groups in the figure: circular binaries are shown in pale orange colour, while eccentric binaries are represented in purple. The first and largest group, which makes up 56\,per cent of the population, consists of circular NS+WD binaries (pale orange) spanning a chirp mass range between 0.35\,M$_\odot$ and 1.2\,M$_\odot$. The formation of circular NS+WD binaries is possible when the NS forms before the WD. In which case, mass-transfer phases following NS formation are likely to spin up (`recycle') the NS, suppress its magnetic field, and circularise the orbit.

We note that within this group there is a small number of binaries ($\sim$0.4\,per cent) with ${\cal M}<0.45\,$M$_\odot$. Such low chirp mass values indicates that the secondary is a Helium-core WD ($M_{\rm WD}<0.5\,$M$_\odot$). These binaries form via a different evolutionary channel compared to those discussed above. The initial binary typically has a large mass ratio and a wide orbit, leading the primary star to fill its Roche lobe as it evolves into a supergiant, resulting in an unstable first mass-transfer phase. Ultimately, the primary star undergoes core collapse, transforming into a NS. Subsequently, when the secondary star fills its Roche lobe, a final stage of mass-transfer is initiated. This phase is typically unstable; however, in instances where it is stable, a low-mass X-ray binary is formed. After the secondary star has lost its hydrogen envelope, it becomes a Helium-core WD. We note that these systems can make up the observed population of millisecond pulsars with low-mass companions. These systems are rare in our models because binaries with less massive companions have less orbital energy and, therefore, are more likely to merge during the CE phase. A recent study by \citet{che21} reports that when using a strong magnetic breaking \citep[e.g.][]{van19} the parameter space for formation of NSs with Helium-core WD companions becomes larger, which could potentially increase the size of this sub-population.

The second group in Fig.~\ref{fig:mfe} comprises binaries with $e\neq0$. These binaries are characterised by chirp masses ${\cal M}>0.8\,$M$_\odot$. The group represents 44\,per cent of the total NS+WD binary population. The formation channel involves the WD forming before the NS. When this happens, the primary star transfers mass to the secondary as it evolves off the main sequence, significantly increasing the secondary's mass. The secondary then becomes more massive than the original primary and ends its life as a NS after the primary has already evolved into a WD. The resulting binary is typically eccentric due to the natal kick received at NS formation (cf. Section~\ref{sec:kicks}).

We note that qualitatively the same picture holds also for $\alpha\alpha2$ and $\gamma\alpha$ model variations. However, in $\alpha\alpha2$ model variations the part of the parameter space with ${\cal M}<0.6$\,M$_\odot$ becomes depleted as these binaries merge before the present time.


\subsection{{\it LISA} detectable population} \label{sec:LISAdetetctable}

The identification of a Galactic GW sources based on the SNR threshold and integration time is not solely determined by the source's intrinsic signal amplitude. Instead, it is also influenced by the specific characteristics of the entire Galactic population, which produces the unresolved Galactic confusion foreground that is added to the {\it LISA} instrumental noise, and against which the source is compared. Therefore, we combine each of our NS+WD population models with respective populations of WD+WD binaries from \citet{kor20}, which have been generated by adopting the same CE models (i.e. $\alpha\alpha$ and $\gamma\alpha$) and the same Milky Way model. As more sources become detectable over time in our data analysis pipeline (cf. Section~\ref{sec:LISApipeline}), they are `removed' from this foreground, thereby reducing its amplitude (and overall noise). The methodology for assessing the detectability with {\it LISA}, as described in Section~\ref{sec:lisa_snr_PE_sec}, enables us to account for these effects. In our analysis we set the nominal mission duration to 4\,yr and the SNR threshold for detectability to $\rho_{\rm 4yr}>7$.

Our simulations reveal subtle yet notable variations in the number of resolved sources, depending on whether the input catalogue includes only WD+WD binaries or both WD+WD and NS+WD binaries. These variations are within a sub-percentage level. When considering WD+WD binaries only, for the $\alpha\alpha$ model, the number of detectable WD+WD binaries increases to $23\times 10^3$ (from $22.9$ reported in Table~\ref{tab:numbers}), representing a variation of approximately 0.52 per cent; for the $\gamma\alpha$ model, the count rises to 24.5$\times10^3$ (from 24.4$\times10^3$), a slight change of about 0.22 per cent. Even variations in the NS+WD population due to different assumptions on natal kicks impact the number of detectable WD+WD binaries, as highlighted in Table~\ref{tab:numbers}. However, we find that the foreground signal remains relatively unchanged when comparing results for WD+WD population and the population that includes both WD+WD and NS+WD binaries. Consequently, we conclude that Galactic WD+WD population constitutes  the major contributor to the unresolved Galactic foreground signal, as previously assumed in the literature.

In Fig.~\ref{fig:hcf}, we display the characteristic strain (i.e. $h_c = \mathcal{A}_n\sqrt{T_{\rm obs}f_n}$ with $T_{\rm obs}=4$\,yr) versus GW frequency for the detectable NS+WD binaries (in colour) in our fiducial model ($\alpha\alpha$-CE +`Verbunt' kick). This population is compared to that of WD+WD binaries (in grey). The black solid line shows the {\it LISA}'s noise level \citep{LISAscireq}, while the grey solid line represents the unresolved Galactic foreground for an assumed 4\,yr observation time, which we compute self-consistently accounting for both NS+WD and WD+WD populations as described in Section~\ref{sec:LISApipeline}. 

For each NS+WD in Fig.~\ref{fig:hcf}, we plot the highest SNR harmonic for binaries with non-zero (true) eccentricity, while for circular binaries we plot the $n=2$ harmonic (the only available one, cf. Section~\ref{sec:lisa_snr_PE_sec}). Our fiducial model's NS+WD detections consist of 62 binaries with $e=0$ and 43 with $e\neq0$; the latter ones are highlighted with a black edge contour. We also find that detectable NS+WD binaries are formed -- i.e., become NS+WD after the second compact object formation -- at frequencies between  $10^{-6}$\,Hz and  $10^{-4}$\,Hz. On average, these binaries are 3.5 Gyr old: circular ones average 1.5 Gyr, while eccentric binaries tend to be around 4 Gyr old. The figure reveals that these two NS+WD sub-populations overlap in the frequency--characteristic strain parameter space, underscoring the importance of eccentricity as a distinguishing measure for inferring their NS+WD formation pathways \citep[see][for similar conclusions regarding NS+NS binaries]{and20,lau20,sto22}.

Across the range of analysed models, we find that in total between $\sim80$ to $\sim350$ binaries reach $\rho_{\rm 4yr}>7$ after 4 years of the mission's lifetime (cf. Table~\ref{tab:numbers}). For comparison, in Fig.~\ref{fig:hcf} we also show an alternative NS+WD population model ($\alpha\alpha2$-CE + `Verbunt' kick). As will be discussed subsequently, this model variation might not accurately represent the Galactic WD+WD population (cf. Section~\ref{sec:discussion}).

As detailed in Section~\ref{sec:lisa_snr_PE_sec}, there are two ways we consider by which we can label a source as having a measurable eccentricity. The first method involves the source having (at least) two harmonics with SNRs above the detection threshold. The number of sources with a measurable eccentricity by this criterion are given in column C of Table \ref{tab:numbers}. The second method involves the minimum detectable eccentricity criterion in Eq.~\eqref{eq:analytic_emin} that was derived in \cite{Mooreetal} based on full Bayesian parameter estimation. The number of sources by this criterion are given in column D of Table \ref{tab:numbers}. To visualise the former result, we colour-code each detectable NS+WD binaries in Fig.~\ref{fig:hcf} by the number of harmonics with $\rho_{\rm 4yr}>7$. 

In our fiducial model, we identify only 5 binaries with multiple detectable harmonics. Hence, relying on the first method (column C) to categorise binaries as eccentric could result in the misclassification of the majority of genuine eccentric sources, increasing the likelihood of confusing these binaries with circular WD+WD and NS+WD binaries. Nevertheless, when applying Eq.~\eqref{eq:analytic_emin}, the majority of the eccentric NS+WD population can be recovered. This observation suggests that while analysing the Galactic population using solely circular waveforms might serve as a good initial data analysis strategy, distinguishing between WD+WD and NS+WD binaries will necessitate eccentric waveforms.

To distinguish between circular NS+WD and WD+WD binaries, it is crucial to measure the binary's chirp mass (via the measurement of $\dot{f}$). For the considered range, the measurement is typically possible at $f \gtrsim2\,$mHz \citep[e.g.][]{kor21}. Figure~\ref{fig:mfe} illustrates that NS+WD chirp masses range between approximately $\sim$0.35\,M$_\odot$ and $\sim$1.2\,M$_\odot$, while the WD+WD chirp mass distribution is expected to peak at around $\sim0.25 - 0.4$M$_\odot$ \citep[depending on the model, e.g. see][]{kor22}, although with a long tail extending up to 1\,M$_\odot$. For frequencies below 2\,mHz, where only GW frequency and amplitude are measured, more massive NS+WD binaries are likely to be indistinguishable from nearby, lower-mass WD+WD binaries. Yet, if the binary's eccentricity can be determined (even in the absence of a chirp mass measurement), any detected nonzero eccentricity would hint at a NS+WD binary. This is because WD+WD binaries in the {\it LISA} frequency range are expected to be circular \citep[e.g.][]{nel01,mar04}, while NS+NS binaries are expected to be an order of magnitude less numerous \citep{LISAastro}.

\subsubsection{Comparing eccentricity distributions}

\begin{figure*}
        \includegraphics[width=0.8\columnwidth]{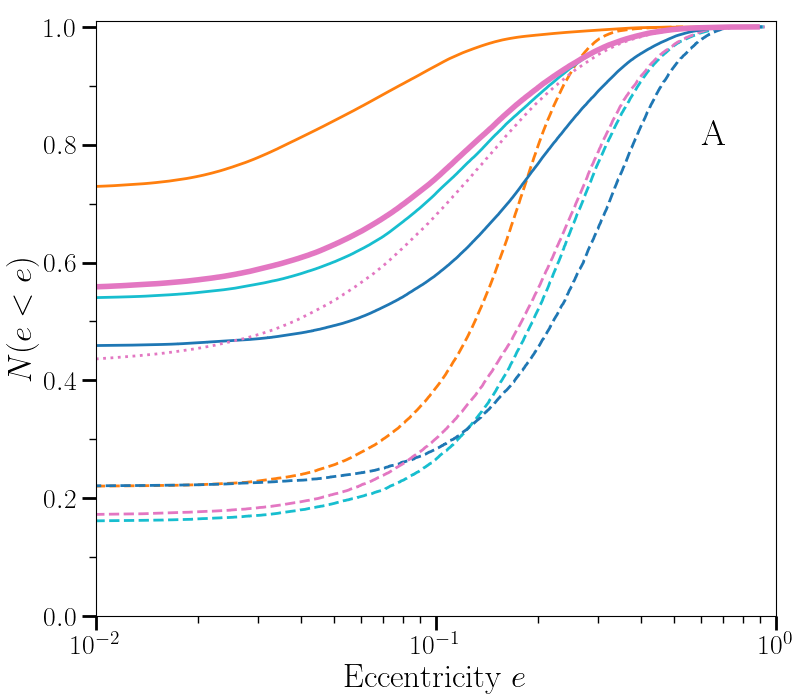}
        \includegraphics[width=0.8\columnwidth]{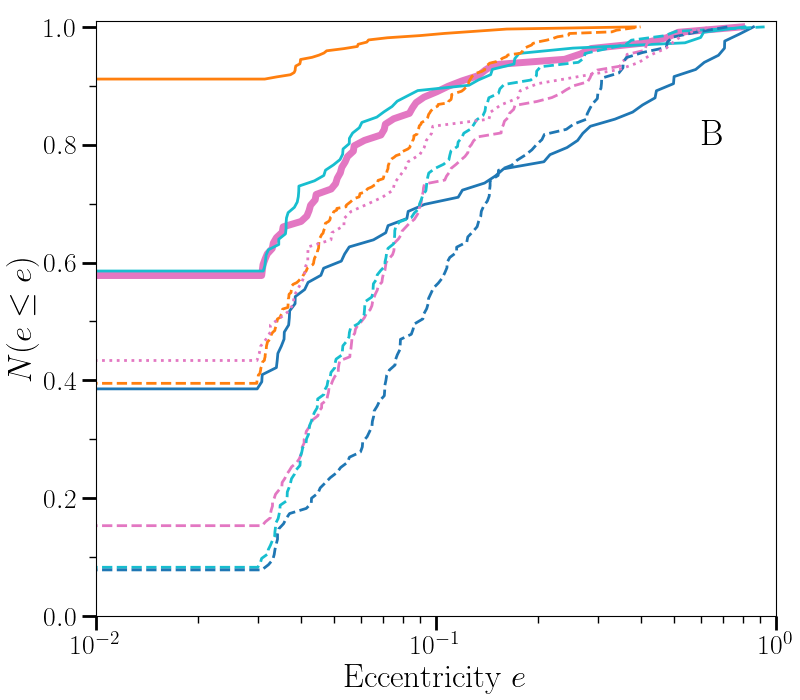}
        \includegraphics[width=0.8\columnwidth]{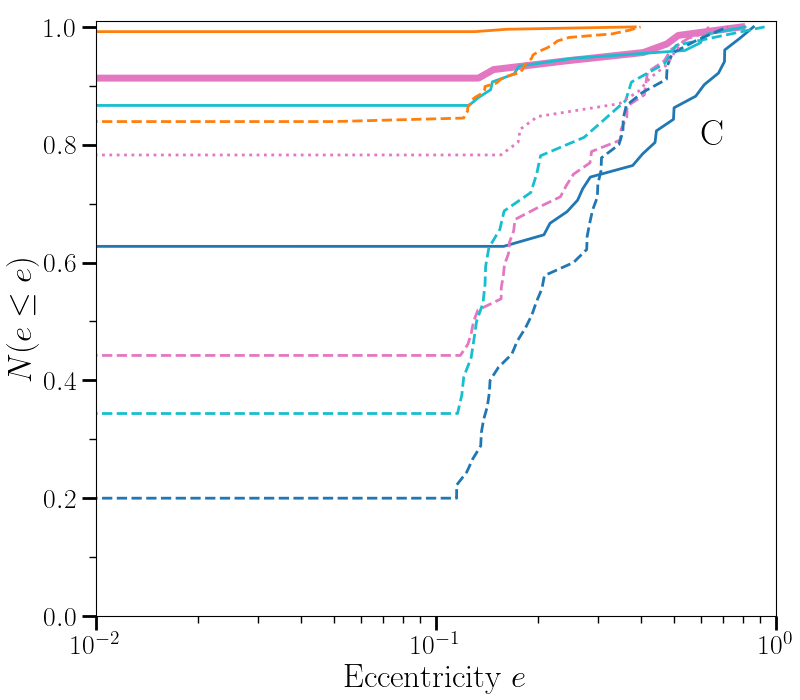}
        \includegraphics[width=0.8\columnwidth]{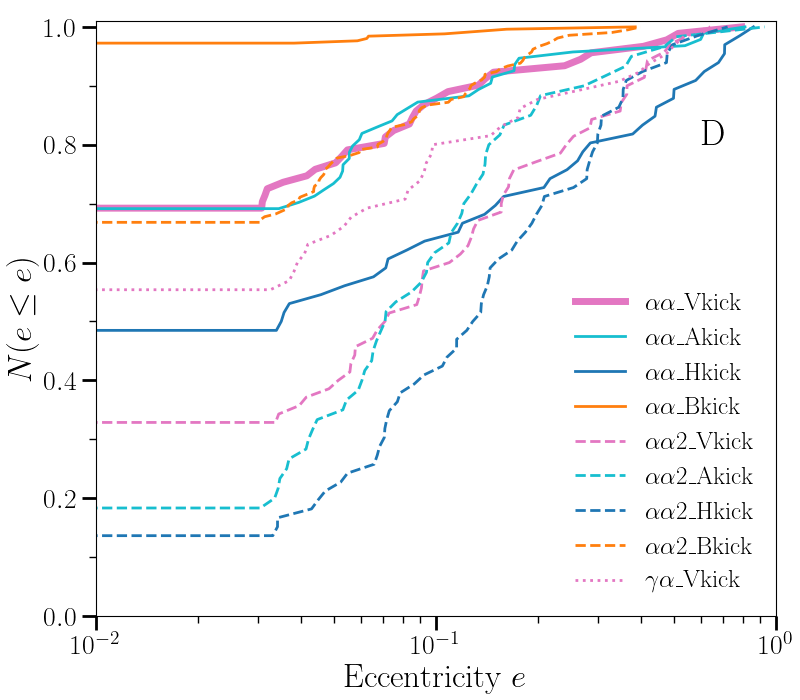}
    \caption{Cumulative eccentricity distributions for NS+WD populations, referenced to Table~\ref{tab:numbers} columns A, B, C and D. Panel A (top left): underlying Galactic NS+WD population within the {\it LISA} band. Panel B (top right): sub-populations with $\rho_{\rm 4yr}>7$. Panel C (bottom left): sub-populations with more than one detectable harmonic. Panel D (bottom right): sub-populations with measurable eccentricities, as assessed using Eq.\eqref{eq:analytic_emin}. Different line styles (solid and dashed) represent models assuming different CE efficiency parameter, while colour variations indicate different natal kick prescriptions, consistent with Fig.~\ref{fig:kicks}. Distributions in panels B, C, and D are normalised to the total number of detectable binaries in their respective models.
    }
   \label{fig:NS+WDecc}
\end{figure*}

In this section, we examine the NS+WD eccentricity distributions across simulated models. Figure~\ref{fig:NS+WDecc} displays cumulative distributions of eccentricities for: the underlying population in the {\it LISA} band (panel A); sub-populations of detectable NS+WD binaries with $\rho_{\rm 4yr}>7$ (panel B); and the same sub-population, but accounting for miss-classification of eccentric binaries as circular, by (re-)assigning $e=0$ to binaries with only one detectable harmonic (panel C); similarly we show sub-populations with measurable eccentricities as confirmed via Eq.~\eqref{eq:analytic_emin} (panel D).

Focusing on the panel A, it is evident that both the CE (highlighted by different line styles) and the natal kick (represented by different line colours) prescriptions play a significant role in shaping the eccentricity distribution. When comparing distributions generated with the same CE efficiency (same line styles), we observe that the `Hobbs' natal kick prescription produces more binaries with $e>0.1$ than the other models. The `Arzoumanian' and `Verbunt' prescriptions yield similar eccentricity distributions, while for the `Blaauw' prescription, only a small fraction of the population (< 90\,per cent) has $e>0.1$. We also observe that in all $\alpha\alpha2$ model variations at least half of the population has $e>0.1$.

Panel B, which displays the same population but with selection effects applied, we observe some changes in the shape of all lines, although they seem to maintain the relative position with respect to each other (with only one exception of $\alpha\alpha2$ CE prescription + `Blaauw' kick).  The median eccentricity shifts to lower values for all models. 

Panel C reveals that when analysing eccentric NS+WD binaries as a collection of circular sources and then counting how many individual binary harmonics reach the detectability threshold (assuming they can be correctly associated with the same source), it becomes evident that many genuinely eccentric binaries can be mislabelled as circular. Furthermore, this criterion appears effective for binaries with an eccentricity larger than 0.1. However, distinguishing between extreme model variations should still be feasible, such as differentiating the $\alpha\alpha$-CE+`Blaauw' kick prescription from the $\alpha\alpha2$-CE + `Arzoumanian'/`Hobbs' natal kick prescriptions.

Lastly, when concentrating on Panel D, it becomes evident that it offers a closer representation of the sub-sample where selection effects are applied (Panel B). However, the median value appears to be biased towards marginally higher values because the eccentricity remains non-measurable for a notable fraction of the population (cf. columns B and D in Table~\ref{tab:numbers}).

\subsubsection{Comparing chirp mass distributions}

 \begin{figure*}
	\includegraphics[width=0.9\columnwidth]{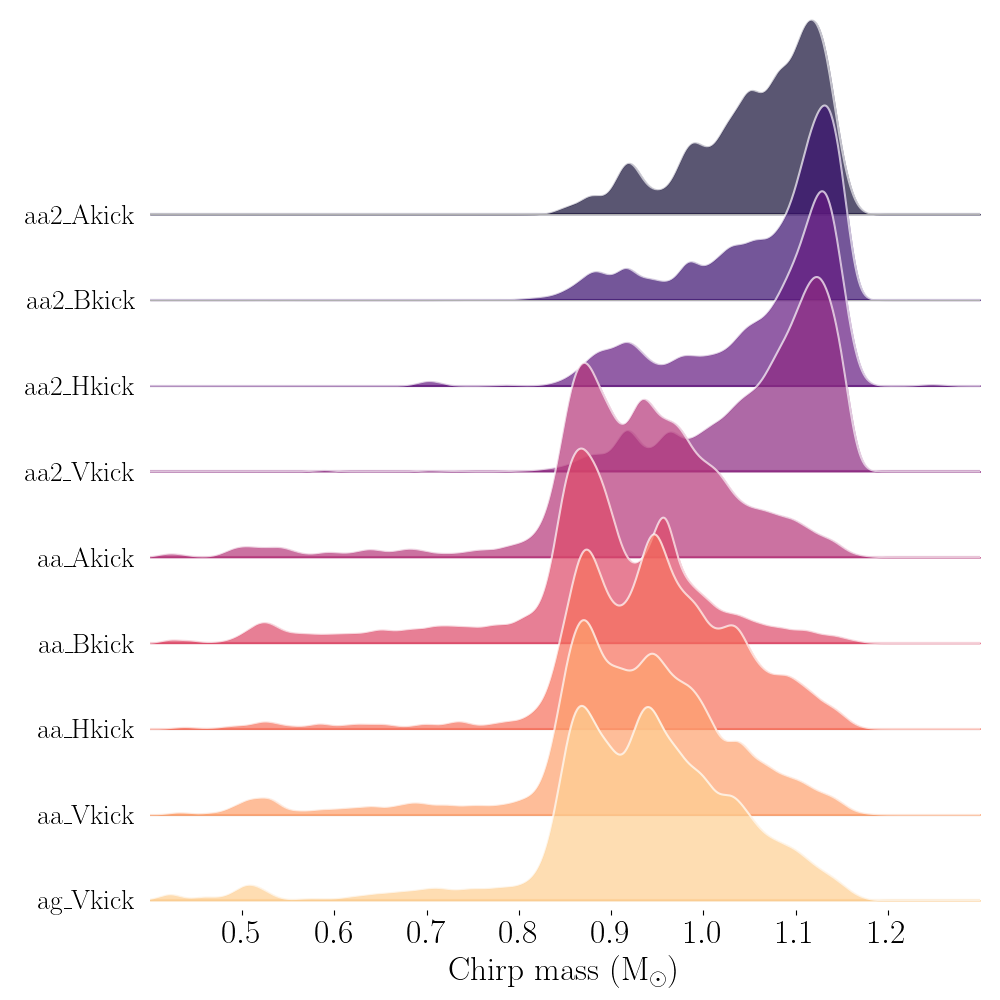}
 	\includegraphics[width=0.9\columnwidth]{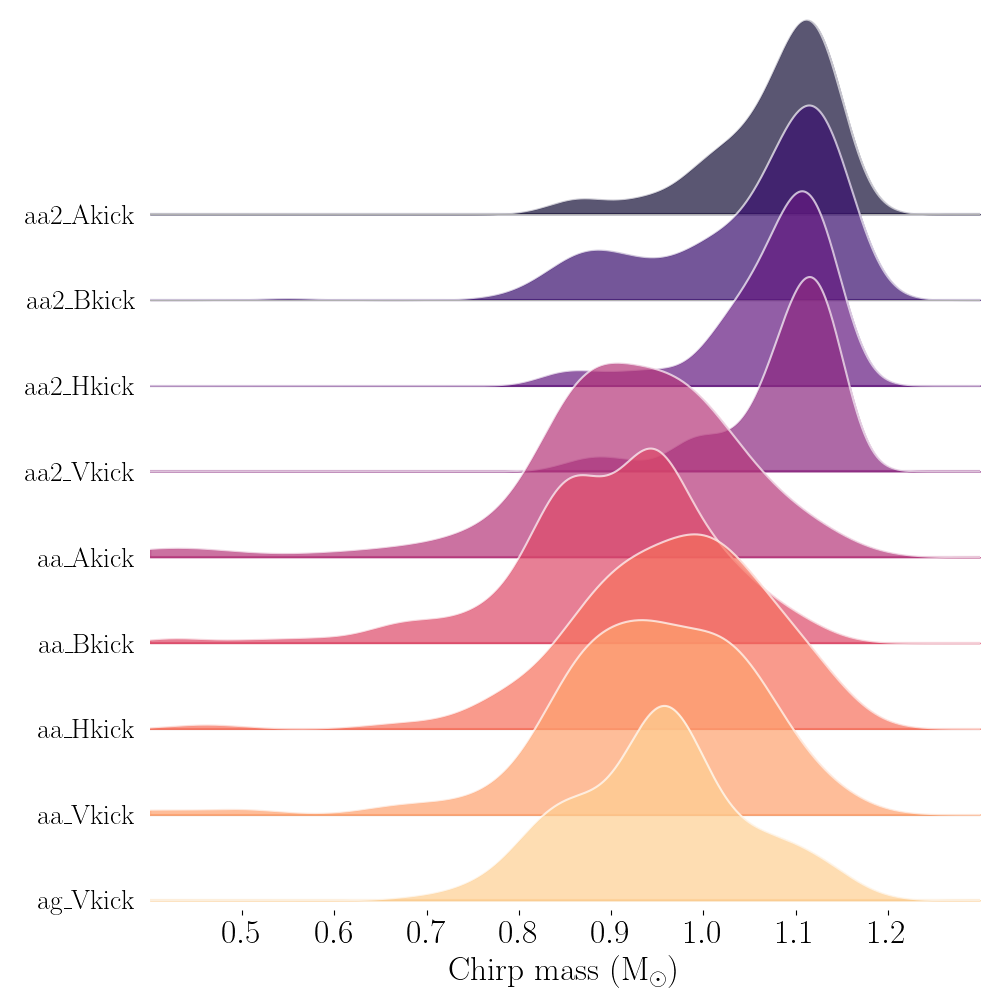}
    \caption{Comparison of the chirp mass density distributions across the suite of considered models. {\it Left panel}: no selection effects; {\it right panel}: including selection effects. }
   \label{fig:mchirp}
\end{figure*}

\begin{figure*}
	\includegraphics[width=0.9\columnwidth]{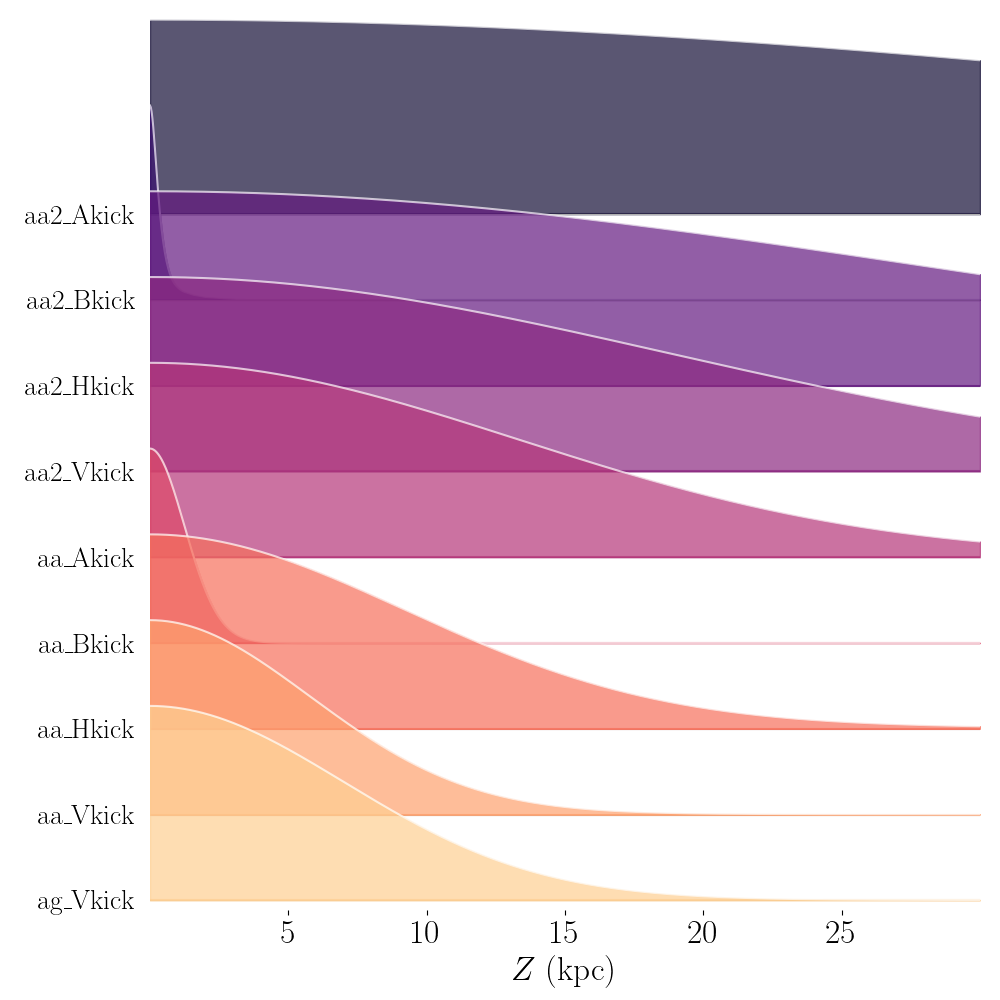}
 	\includegraphics[width=0.9\columnwidth]{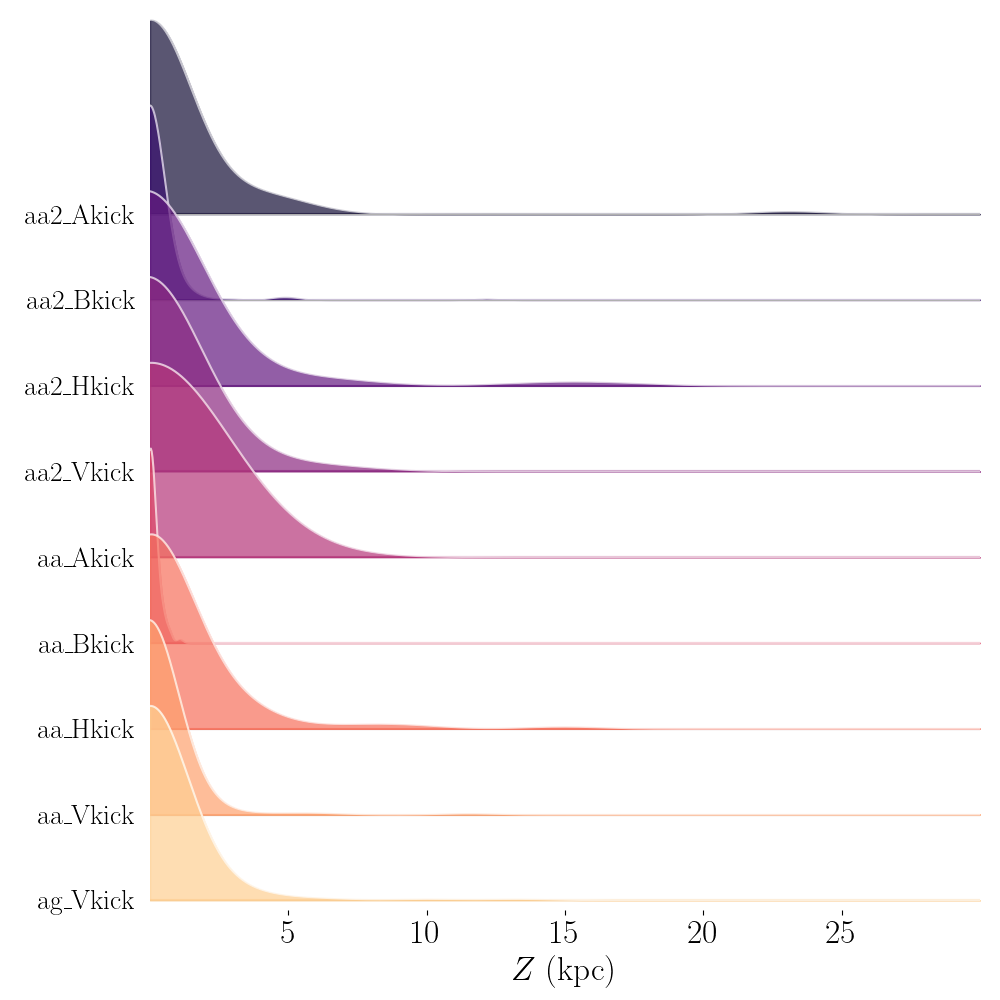}
    \caption{Comparison of NS+WD distribution in $Z$ coordinate across the suite of considered models. {\it Left panel}: no selection effects; {\it right panel}: including selection effects.}
   \label{fig:Z}
\end{figure*}

As previously discussed, chirp mass measurement is essential not only for characterising NS+WD binaries but also for distinguishing them from other types of Galactic double compact objects, particularly when NS+WD chirp mass distributions may overlap with those of WD+WD and/or NS+NS binaries. In Fig.~\ref{fig:mchirp}, we compare chirp mass probability densities across the considered NS+WD models: the left panel displays the underlying population (i.e. no selection effects), the right panel shows binaries with $\rho_{\rm 4yr}>7$.  

Firstly, we observe that in all models, NS+WD chirp masses range between 0.4\,M$_\odot$ and 1.2\,M$_\odot$, peaking either at $\sim0.9$\,M$_\odot$ (for $\alpha\alpha$ and $\gamma\alpha$ model variations) or at $\sim1.1$\,M$_\odot$ (for $\alpha\alpha2$ model variations). Thus, if the chirp mass is constrained to better than 30\,per cent, NS+WDs can be separated from WD+WDs, which typically have lower chirp masses. The reason for larger NS+WD chirp masses is twofold: the presence of the more massive NS component ($1.2-1.35$\,M$_\odot$) and the typically more massive WD component ($0.95$\,M$_\odot$ on average). 

Distinguishing NS+WD from NS+NS binaries based on the chirp mass will be more challenging, especially given that the number of NS+NS binaries is expected to be in the tens, while NS+WD binaries are anticipated to be in the hundreds \citep{LISAastro}. For example, models from \citet{vig20} and Fig.~3 of \citet{kor21} indicate that Galactic NS+NS chirp masses may peak at $\sim1.1$\,M$_\odot$. This implies that a chirp mass constraint of better than 10\,per cent is necessary to distinguish NS+NSs from NS+WDs in our $\alpha\alpha$ and $\gamma\alpha$ models, while the majority of NS+WDs in our $\alpha\alpha2$ models may be indistinguishable from NS+NSs (based on the chirp mass only).
 
Focusing on the left panel of Fig.~\ref{fig:mchirp}, we observe a significant difference between chirp mass distributions in $\alpha\alpha2$ models compared to $\alpha\alpha$/$\gamma\alpha$ models. It is important to recall that the difference between these models stems from the varying assumptions for the CE efficiency (cf. Section~\ref{sec:CE}): our $\alpha\alpha2$ models are set to be less efficient in expelling the envelope, resulting in a longer CE phase, a more significant shrinkage of the binary orbit, and a higher overall number of binaries merging before the present time. Thus, if the NS+WD population can be accurately isolated from that of WD+WD, the chirp mass distribution can provide a link for to the CE efficiency parameter.

By comparing the two panels of Fig.~\ref{fig:mchirp}, we find that the main feature of the chirp mass distributions -- i.e. the location of the peak -- are preserved across all model variations; although sub-structures are lost due to the reduced number of binaries. 
Therefore, we suggest that the position of the peak of the NS+WD chirp mass distribution can be used to distinguish between model variations assuming different CE prescriptions.

\subsubsection{Comparing $Z$ distributions} \label{sec:Zdist}

Lastly, we investigate how the positions of NS+WD binaries in our mock Milky Way are influenced by the NS natal kicks (cf. Fig.~\ref{fig:kicks}). For example, \citet{rep17} demonstrated that the root mean square of the height above the Galactic plane for black holes and NSs in X-ray binaries can be used as a proxy to discriminate among different natal kick prescriptions. Following the same idea, we consider the positions of NS+WD binaries as measured by {\it LISA}. In Fig.~\ref{fig:Z}, we compare the $Z$ distributions across the considered NS+WD models: on the left, no selection effects are applied, while on the right we show binaries with $\rho_{\rm 4yr}>7$.

We observe that in the top panel, both the NS natal kick prescription and the CE efficiency effects are noticeable. The CE efficiency determines the binary orbital separation at NS+WD formation and, consequently, the binary orbital energy. Depending on the magnitude of the natal kick, the binary may or may not be disrupted. Thus, both the CE  and the natal kick magnitude play a role in determining the position of a binary in the Galaxy at the present day. For a fixed CE model, the `Arzoumanian' model produces the broadest $Z$-distribution, followed by the `Hobbs' and `Verbunt' models; the `Blaauw' prescription results in a significantly narrower $Z$-distribution. 
When {\it LISA} selection effects are applied, these differences become less pronounced because only tightest, highest frequency binaries can be detected at $Z$ larger than several kpc. Still, models using the `Blaauw' prescription remain visually narrower compared to the rest (see also Fig.~\ref{fig:kicks}).

\section{Discussion}\label{sec:discussion}

In this study we assembled a suite of NS+WD populations models using the binary population synthesis technique. We assessed the size and the properties of the {\it LISA} detectable population and investigated the ways NS+WDs can be differentiated from WD+WDs in the data based on the astrophysical properties of the population: chirp mass and eccentricity distributions as well as binaries' 3D positions in the Galaxy. In particular, distinguishing between the binary types is not trivial as both WD+WD and NS+WD binaries will appear in the {\it LISA} data as quasi-monochromatic sources and will occupy a similar parameters space (cf. Fig.~\ref{fig:hcf}). Therefore, it will require a combination of GW measurement (such as chirp mass, eccentricity and Galactic position) measurement and electromagnetic follow  up (e.g. looking for pulsar signatures). 

As the primary objective of this study is to provide guidelines for the analysis and interpretation of future {\it LISA} data, rather than fitting models to the observed sample, some of the proposed model variations may not be suitable for describing both NS+WD and WD+WD populations. For instance, the assumption regarding the efficiency parameter $\alpha$ in CE interactions (incorporated into our binary population synthesis modelling as the product $\alpha\lambda$) plays a pivotal role in determining the number of detectable binaries (cf. Table~\ref{tab:numbers} and Fig.~\ref{fig:hcf}). While this assumption appears to align well with studies focused on post-CE binaries \citep{zor10, zor14, too13}, it does not accurately replicate the observed WD+WD population \citep{too12,too17}. However, a recent study by \citet{sch23}, which performs reverse modelling of WD+WD binaries' formation histories, also derived estimates of lower CE efficiency than those assumed in our fiducial model. Furthermore, it is interesting to observe that $\alpha\alpha2$ set of model variations does not introduce any confusion foreground (cf. Fig.~\ref{fig:hcf}). An observationally-driven WD+WD population model that incorporates distributions based on observed spectroscopic samples of WD+WD candidates produces the confusion foreground of a comparable amplitude to our fiducial model \citep[see][]{kor22}.

We acknowledge that our Milky Way model may be somewhat simplistic. For example, we represent the stellar Galactic disk as a single component, rather than differentiating between the thin and thick disks. Nevertheless, we have incorporated a total stellar mass and a star formation history that encompass both the thick and thin disk stellar populations. While we have not conducted model variations to specifically explore the impact of our Galactic model on the {\it LISA} detectable NS+WD population, relevant studies for NS+NS binaries by \citet{sto22} and for WD+WD binaries by \citet{geo23} have indicated that changes in birth positions—such as those resulting from implementing two-disk components with distinct scale parameters—do not significantly affect the resolvable frequency or eccentricity of these binaries. Furthermore, \citet{sto22} have found that detailed integration of Galactic orbits for NS+NS binaries is not crucial for accurately predicting the LISA-resolved NS+NS population. However, they caution that the choice of star formation history is critical when predicting LISA-visible NS+NS populations. This consideration may also be pertinent to NS+WD binaries, especially given their potentially shorter formation timescale compared to WD+WD binaries.

We also recognise that our models only captures NS+WD binaries forming in the field and do not include those binaries that may form in globular clusters. Given that globular clusters are known to host a substantial fraction of the ultra-compact X-ray binary population, it is reasonable to expect a considerable number of NS+WD binaries in these clusters. Notably, the formation rate of low-mass X-ray binaries (LMXBs), which could potentially include NS+WD binaries, is estimated to be considerably higher per unit mass in globular clusters than in the Galactic field \citep[e.g.,][]{hei03,pad23}. \citet{kre18}, for example, predicts the existence of approximately $10^3$ mHz NS+WD binaries in the Milky Way's globular clusters, with a few to several of these potentially detectable by {\it LISA}.

\subsection{Prospects for measuring NS+WD total mass via periastron advance}

\citet{set01} pointed out that for eccentric binaries, a measurement of the periastron advance -- analogous to the perihelion advance of Mercury -- may be detectable in binaries in the {\it LISA} data. 
In the absence of periastron precession, the GW radiation is emitted at {\it exactly} multiples of the orbital frequency \citep{pet64}.
If the binary is precessing, the spacing of the individual harmonics will be shifted by integer multiples of the periastron precession frequency
\begin{align} \label{eq:periastron_prec_freq}
    f_{\rm PP} &= \frac{1}{1-e^2}\left(\frac{M}{{\rm M}_\odot}\right)^{2/3}\left(\frac{f_{\rm GW}}{1\,\mathrm{mHz}}\right)^{5/3} \times 0.29\,\mathrm{year}^{-1}.
\end{align}
From the above equation it is clear $f_{\rm PP}$ depends on the total mass $M$ of the binary, while $\dot{f}_n$ depends on the chirp mass (cf. Eq.~\ref{eq:fdot}); \cite{Mooreetal} illustrated how these two measurements can be combined to derive the individual component masses.
It seems likely that for the majority of binaries where the eccentricity can be measured, it will be possible to obtain the measurement of the component masses by this technique.
While this will depend on the binary's SNR and frequency, we can speculate that results reported in column D of Table~\ref{tab:numbers} are representative of the number of individual component mass measurements. 
In the example investigated by \citet[][see their fig.~4]{Mooreetal}, the fractional uncertainty on the primary mass (90\,per cent uncertainty) is $\sim$33\,per cent.

For completeness, we highlight that \citet{tau18} discussed a method for measuring NS masses in NS+WD binaries leveraging a well-established correlation between the orbital period and the mass of a Helium-core WD in LMXB systems. Only those post-LMXB NS+WD binaries with less than about 9-hour orbital periods, which can coalesce within a Hubble time to become visible LISA sources, are observed to have WD masses within a remarkably narrow range ($M_{\rm WD} = 0.162 \pm 0.005$\,M$_\odot$). This fact in combination with {\it LISA}’s chirp mass measurement facilitates an accurate determination of NS masses with a precision within estimated precision of $\sim4$\,per cent (see their figure 4).

\subsection{Systemic velocities of millisecond radio pulsars}
\label{s:syst_velocities}

\begin{figure}
        \includegraphics[width=\columnwidth]{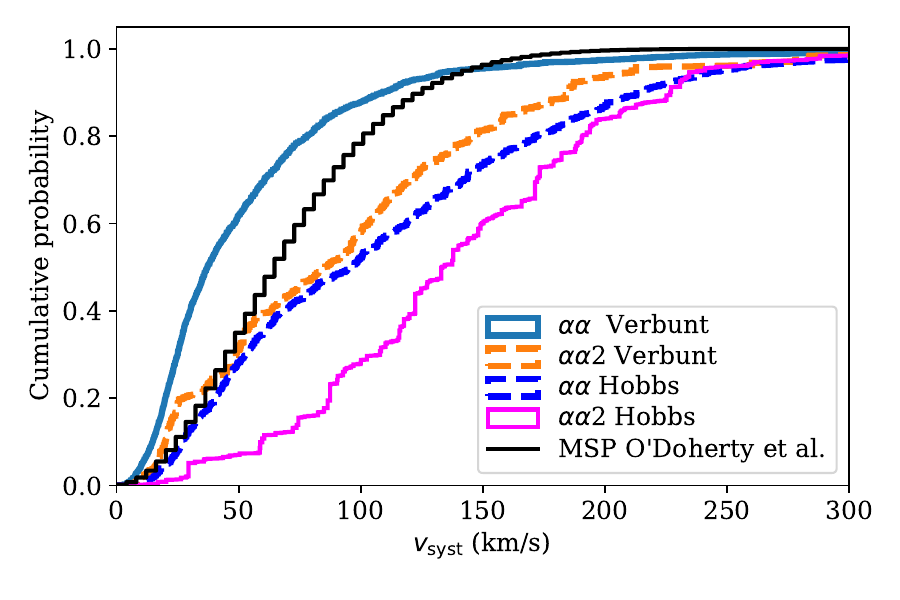}
    \caption{Distribution of NS+WD systemic velocities  for binaries with small eccentricities $e < 0.1$.}
   \label{fig:syst_vel_verb}
\end{figure}

Recently \cite{ODoherty2023arXiv} performed analysis of systemic velocities for binaries with NS including millisecond radio pulsars. They integrated orbits of systems back in time and recorded their velocities at the expected moment of compact object formation. These velocities are typically smaller than the NS natal kick and ranges from tens to $\approx 250$~km~s$^{-1}$. Here we perform simple comparison between systemic velocities of NS+WD binaries with small eccentricities ($e < 0.1$) and their results. For this comparison, we select all binaries which satisfy only single aforementioned criteria i.e. systems plotted are neither selected on the base of the LISA detectability nor their radio detectability as radio pulsars. We show the result of our analysis in Figure~\ref{fig:syst_vel_verb}. The curve obtained by \cite{ODoherty2023arXiv} is between $\alpha\alpha$ and $\alpha\alpha$2 models with `Verbunt' natal kicks. The $\alpha\alpha$ model with Hobbs natal kick provides similar results as $\alpha\alpha$2 `Verbunt' natal kicks. The small systemic velocities ($v_\mathrm{syst} < 60$~km~s$^{-1}$) follows $\alpha\alpha$2+`Verbunt' (similarly $\alpha\alpha$ +  `Hobbs') cumulative distribution, but later on the curve suggested by \cite{ODoherty2023arXiv} diverges from these curves strongly and it follows $\alpha\alpha$+`Verbunt' model quite closely for $v_\mathrm{syst}>120$~km~s$^{-1}$.  

It is interesting to note that estimates of \cite{ODoherty2023arXiv} fall between two curves produced with the same natal kick prescription but with different assumptions about the CE evolution. This suggests that  values of $\alpha\lambda$ for millisecond pulsar formation should be between 0.25 and 2. We can also speculate that the expected \textit{LISA} population would be more similar to our simulations with the `Verbunt' natal kick prescription.

\subsection{Follow up radio observations}

Some of NS+WD binaries discovered by {\it LISA} could host millisecond or normal radio pulsars. It is much more probable that follow-up radio observations discover a new millisecond radio pulsar than a normal radio pulsar because radio emission from normal pulsar is beamed much stronger than in the case of millisecond radio pulsar \citep[e.g.][]{Kramer1998ApJ}. Having discovered NS+WD {\it LISA} source, it is possible to design an optimal strategy for radio follow-up. {\it LISA}'s sky localisation for Galactic stellar-remnant binaries largely depends on the binary's SNR and frequency, and is expect to be in the range of several to few deg$^2$ at $f_{\rm GW}>2-3$\,mHz \citep[e.g.][]{fin22, Mooreetal, LISA2Redbook}. A typical pulsar survey covers $1.5^\circ$ in a single pointing like the Parkes multi-beam pulsar survey \citep{Manchester2001MNRAS}. Thus, if NS+WD system is subsequently discovered in a dedicated pulsar search, radio observations will enormously improve the sky localisation because pulsar coordinates are typically known with accuracy exceeding 1~arcsec. Pulsar discovery will also help to constrain the orbital parameters such as the mass function and eccentricity via the pulsar timing. 

The normal radio pulsar could be formed if the NS is formed after WD formation via the reversed channel. Additionally, the system should not be too old because normal pulsars stop operating typically after 0.1-1~Gyrs. Thus, highly eccentric systems with NSs can be the primary targets for dedicated young pulsar searches. The expected pulsar orbital periods will be ranging from a few tens of milliseconds to a couple of seconds, see \cite{Igoshev2022MNRAS} for recent estimates on initial NS spin periods. Only a fraction of about $10$~per cent of all young pulsars in NS+WD can be detected \citep{TaurisManchester1998MNRAS} because in multiple cases the beam will miss the Earth due to the geometric orientation of the NS. Thus, given the number counts in Table~\ref{tab:numbers} and age distribution we could expect to discover $\approx 2-6$ young pulsars in a dedicated radio search if an efficient technique to deal with orbital accelerations is implemented. The lower estimate corresponds to the pulsar lifetime of 0.1~Gyr while upper one corresponds to lifetime of 1~Gyr. Additional constrains might arise from the limited sky coverage of radio surveys. Thus, the Parkes Multibeam Pulsar Survey \citep{Manchester2001MNRAS} covered only the Galactic longitude $|b| < 5^\circ$. So, these radio pulsars could be missed in radio surveys if the relevant part of the sky is not covered.

The millisecond radio pulsar is typically formed as a result of stable mass-transfer from secondary to NS. Thus, we expect that millisecond pulsar candidates have nearly circular orbit with eccentricity well below $0.1$. The mass is transferred together with angular momentum which spins up the NS and it typically has a period shorter than $\approx$30~milliseconds. The millisecond radio pulsars could operate for many Gyr and their average radio profiles are much wider than that is for normal radio pulsars \citep{Kramer1998ApJ}. The profile width reaches 100$^\circ$. This opening angle could correspond to up to 50~per cent possible detection. Thus, many more (up to $\approx$ 50) are expected to be discovered in radio follow-ups among NS+WD GW emitters with small eccentricity. The caveat is that radio signal from millisecond pulsars located in regions with high free electron density (e.g. the Galactic Centre) could be smeared and leave these pulsars undetectable.

\section{Summary and conclusions}

In this study, we examine the detectability of Galactic NS+WD binaries with {\it LISA}, estimating a population size of $\mathcal{O}(10^2$) detectable systems. According to our fiducial binary population synthesis model about a half of the NS+WD population in the {\it LISA} band are circular. These are formed through the reversed channel (i.e. WD forms first, followed by the NS; see also \citealt{he24} on how this result may change when changing assumptions in the underlying binary evolution model). The remaining population consists of eccentric NS+WD binaries that originate from the direct channel, where the NS forms first, followed by the WD. In the current study, we have not undertaken a full reconstruction of the population from the simulated {\it LISA} data. Instead, our investigation has concentrated on exploring how NS+WD binaries can be differentiated from other types of Galactic binaries within the LISA dataset, as well as on identifying key observable features that are essential for aiding data inference. Given the high completeness of {\it LISA}'s sample up to frequencies of 2-3\,mHz \citep{LISAastro}, we anticipate that it will be possible to recover the underlying true formation rates and population properties from LISA data, contingent upon the identification of NS-WD binaries within the dataset.

The identification of NS+WD binaries within the {\it LISA} data may be challenging, especially at frequencies below 2\,mHz, where they might be misclassified as WD+WD binaries. In particular, this misclassification risk is heightened for circular NS+WD binaries, which can be easily confused with WD+WDs when the chirp mass measurement is absent (typically at frequencies lower than a few mHz). However, eccentric NS+WD binaries can potentially be pinpointed through the eccentricity measurement. We explored two methods to identify sources with measurable eccentricity in our catalogues. The first requires a source to present at least two harmonics above the detection threshold, mimicking the analysis of the Galactic population using only circular waveforms. In contrast, the second leverages the minimum detectable eccentricity criterion, as derived in \citet{Mooreetal} using eccentric waveforms for source parameters recovery. Relying exclusively on the first method carries a risk of misidentifying the majority of genuine eccentric sources. However, the second method, which employs the minimum detectable eccentricity criterion, proves more adept at distinguishing eccentric sources. While initial analyses — particularly within the {\it LISA} {\it Global fit} framework \citep[e.g.][]{lit23} — might rely on circular waveforms, a more clear differentiation between WD+WD and NS+WD binaries calls for the inclusion of eccentric waveforms \citep[for a detailed analysis, refer to][]{Mooreetal}. We aim to further investigate source confusion through a more realistic {\it Global fit} type of analysis in a future study.

We have also demonstrated that the NS+WD chirp mass distribution is sensitive to the CE efficiency parameter by comparing $\alpha\alpha$ and $\alpha\alpha2$ sub-sets of models. In particular, we propose the idea that the position of the peak of the chirp mass distribution can be used to distinguish between these model variations. Additionally, our analysis underscores the sensitivity of the eccentricity distribution to the NS natal kick prescription, emphasising the potential to distinguish (at least) between extreme kick prescriptions based on the subset overall population accessible with {\it LISA}. However, learning about the NS natal kicks based on NS+WD positions above the Galactic plane -- as it has been done for high-mass X-ray binaries in \citet{rep17} -- might be  challenging. This is due to the constraints on position measurements with {\it LISA}, a marked difference when compared to high-mass X-ray binaries, where measurements are typically more precise (deg$^2$ versus arcsec$^2$).

Lastly, we discussed the prospects for identifying electromagnetic counterparts of NS+WD binaries in the radio regime, which could offer additional insight into the formation channels and characteristics of these systems. Our preliminary calculations suggest the possibility of detecting a few young pulsars with WD companions and several tens of millisecond pulsars through targeted radio follow-up.

In conclusion, our exploration of the detectability of Galactic NS+WD binaries with the upcoming {\it LISA} mission has highlighted the potential of unlocking of NS-specific science with the {\it LISA} data. We have pinpointed challenges in classifying NS+WD binaries accurately, owing to the abundance of WD+WD binaries and the potential overlap with NS+NS binaries. Our conclusions, together with those in the companion paper by \citet{Mooreetal}, carry significant implications for defining {\it LISA} data analysis strategies and data interpretation. 

\section*{Acknowledgements}
We are grateful to the referee for their careful reading of the manuscript and insightful comments. We also thank the {\it LISA} Birmingham group, particularly Hannah Middleton, Riccardo Buscicchio, Diganta Bandopadhyay, and Alberto Vecchio, for their fruitful discussions during group meetings. We also wish to acknowledge that preliminary work on this topic began with Ms. Abbie Nicholls' MSc thesis, a (2019/2020) Physics graduate from the University of Birmingham.\\
VK and ST acknowledge support from the Netherlands Research Council NWO (respectively Rubicon 019.183EN.015, and VENI 639.041.645, VIDI 203.061 grants).
Work of A.P.I. is supported by STFC grant no.\ ST/W000873/1.\\
This research made use of the tools provided by the \textit{LISA} Data Processing Group (LDPG) and the \textit{LISA} Consortium \textit{LISA} Data Challenges (LDC) working group\footnote{\href{https://lisa-ldc.lal.in2p3.fr/}{https://lisa-ldc.lal.in2p3.fr/}}.\\
This research made use of \textsc{astropy}\footnote{\href{http://www.astropy.org}{http://www.astropy.org}}, a community-developed core \textsc{python} package for Astronomy \citep{Astropy_2013, Astropy_2018}, \textsc{matplotlib} \citep{Hunter_2007}, \textsc{numpy} \citep{Numpy_2006, Numpy_2011}, \textsc{scipy} \citep{Virtanen_2020}, as well as \textsc{galpy}\footnote{\href{http://github.com/jobovy/galpy}{http://github.com/jobovy/galpy}} \citep{Bovy2015ApJS} and \textsc{joypy}\footnote{\href{https://github.com/leotac/joypy}{https://github.com/leotac/joypy}} packages.

\section*{Data Availability}

Mock catalogues produced as part of this study are available on Zenodo \href{https://doi.org/10.5281/zenodo.10854469}{10.5281/zenodo.10854469}.


\bibliographystyle{mnras}
\bibliography{NSWD} 






\bsp	
\label{lastpage}
\end{document}